\documentclass[11pt]{article}
\pdfoutput=1 
\usepackage{jheppub} 

\usepackage[T1]{fontenc}
\usepackage{xcolor}
\usepackage{caption}
\usepackage{subcaption,longtable,stmaryrd}
\usepackage{bigints}
\usepackage{multicol}
\usepackage{tikz,lipsum,lmodern}
\usepackage{dsfont}
\usepackage[most]{tcolorbox}
\usepackage{blindtext}
\usepackage{adjustbox}
\title{Aspects of $T\Bar{T}+J\Bar{T }$ deformed Schwarzian: From gravity partition function to late-time spectral form factor}
\author[]{Arpan Bhattacharyya,}
\author[]{Saptaswa Ghosh}
\author[]{and Sounak Pal
}
\affiliation[]{\it Indian Institute of Technology, Gandhinagar,\\
Gujarat-382055, India}

\emailAdd{abhattacharyya@iitgn.ac.in}
\emailAdd{saptaswaghosh@iitgn.ac.in}
\emailAdd{palsounak@iitgn.ac.in}
\abstract{In this paper, we investigate different thermodynamic properties of $T\Bar{T}+J\Bar{T}$ deformed Schwarzian theory and its different gravitational perspectives. First, we compute the partition function of $U(1)$ coupled 2D-gravity with fixed chemical potential, obtained from the dimensional reduction of the four-dimensional Einstein-Maxwell theory. Then, we compute the partition function of the  gravity theory which is the dual to the deformed Schwarzian living on its boundary and study the genus expansion of the one and two-point correlation function of the partition function of the theory. Subsequently, we use the one-point function to compute the ``Annealed'' and ``Quenched'' free energy in low-temperature limits and make a qualitative comparison with the undeformed theory. Then, using the two-point function, we compute the Spectral Form Factor of the deformed theory in early and late time. We find a dip and ramp structure in early and late time, respectively. We also get a plateau structure in the $\tau$-scaling limit. Last but not least, we comment on the late-time topology change to give a physical interpretation of the ramp of the Spectral Form Factor for our theory.
}
\begin{document}
\maketitle
%%%%%%%%%%%%%%%%%%%%%%%%%%%%%%%%%%%%%%%%%%%%%%%%%
\section{Introduction}\label{intro}
Einstein's theory of General Relativity has been a remarkable success in describing the classical theory of gravity. However, the theory conflicts with the laws of quantum mechanics. Among the four fundamental forces, quantizing gravity remains fundamental physics's most difficult and unresolved problem. For instance, in the path integral formalism of quantum field theory, we need to integrate over all field configurations. When applying the formalism in the case of gravity, one has to integrate over all possible metric field tensor ($g_{\mu\nu}$) governed by the Einstein-Hilbert Action. However, the classical understanding of spacetime is not very rigorous due to the intricacy of the structure, such as topology, signatures, etc. Due to those facts, studying quantum gravity in four spacetime dimensions is mathematically (as well as physically) challenging. Due to the facts discussed above, it would be better to study the lower dimensional models of quantum gravity to get physical and mathematical intuition.\par 
In $2D$, the Euclidean Einstein-Hilbert action can be integrated out using the Gauss-Bonnet theorem and gives a topological term$\sim e^{-S_{0}\chi}$, where $\chi$ is the Euler characteristic and $S_{0}\sim\frac{1}{G_N}$. In recent days, the models of 2D quantum gravity \cite{Almheiri:2014cka}, especially the Jackiw-Titelboim (JT) gravity \cite{Jackiw:1984je, Teitelboim:1983ux}\footnote{For more details, interested readers are referred to  \cite{Mertens:2022irh, Moitra:2019bub, Moitra:2022glw}.}, a $2D$ Euclidean model of dilaton gravity on spacetime with a negative cosmological constant, has gained much attention. JT gravity exhibits many interesting features. For example, it has been shown that the partition function of JT gravity has two main ingredients: one is the volume of the moduli spaces, and another is the asymptotic $AdS_{2}$ boundary. \textcolor{black}{In the context of $AdS/CFT$ correspondence \cite{Maldacena:1997re, Witten:1998qj}, the latter has a holographic counterpart: one-dimensional Schwarzian quantum mechanics \cite{Maldacena:2016upp}.} From this holographic perspective, one can also compute the disk partition function, which is one of the building blocks of studying the genus expansion of the partition function. Apart from this, the computations of higher genus partition functions have revealed the relation with the random matrix model (RMT)\cite{Saad:2019lba, Stanford:2019vob}. Recently, some studies on JT gravity and its connection to Karch-Randall Braneworld have been done in \cite{Geng:2022tfc, Geng:2022slq}\footnote{For more applications, interested readers are referred to, e.g. \cite{Bhattacharya:2023drv, Aguilar-Gutierrez:2023tic}.}.  \par 
Additionally, several recent studies have been performed to understand the non-perturbative effects of Newton's constant $G_N$ in the context of JT-gravity \cite{Johnson:2021tnl}. The main point of these investigations is to show that non-perturbative saddles appear in the gravitational path integral late times, which are typically irrelevant at early times.  Also, the energy eigenvalues for chaotic systems usually repel and are described by the Sine Kernel \cite{Okuyama:2023pio,Anegawa:2023klh}. Eventually, the chaotic property of the underlying theory can be directly investigated by studying the behaviour of the \textit{Spectral Form Factor (SFF)}, which is defined by the two-point correlator (analytically continued) of the partition function: $\Big\langle Z(\beta+it)Z(\beta-it)\Big\rangle$. \textcolor{black}{ In JT gravity, the above contribution comes from the double-cone cylindrical topology at the leading order.}\footnote{In principle, one can compute this two-point function by summing over an arbitrary genus. But it is computationally more involved.} In general, for a chaotic quantum system such as the Sachdev–Ye–Kitaev model(SYK) model, the $SFF$ consists of three regions: dip in early time, ramp, and the constant region (plateau) in late time. Interestingly, the low energy dynamics of the SYK model \cite{1993PhRvL..70.3339S, Kitaev:2017awl,syk, Maldacena:2016hyu, Sarosi:2017ykf} can be described by the one-dimensional Schwarzian quantum mechanics. So, it is implicit that the $SFF$ for JT-gravity also shows the same nature as the SYK model %\textcolor{red}{though it is still an open question whether black holes correspond to a \textit{single random matrix model} or not}. 
The first kick-off study of the Spectral Form Factor (for theories including the SYK model relevant for holography) has been done in \cite{Garcia-Garcia:2016mno, Papadodimas:2015xma,  Cotler:2016fpe}. Apart from those, there have been many studies investigating the spectral properties for (JT) gravity, random matrix models, spin chains, and conformal field theories \cite{Anegawa:2023klh, Okuyama:2023pio, Castro:2023rfd, DiFrancesco:1993cyw, Mukherjee:2020sbt, PhysRevD.100.026017, He:2022ryk, Caceres:2022kyr, Khramtsov:2020bvs, Winer:2022ciz, Winer:2023btb, Zhou:2023qmk, Bhattacharyya:2023zda, Balasubramanian:2016ids, Dyer:2016pou, Krishnan:2016bvg}.  \par
The thermodynamics of JT gravity has been investigated by summing over all possible geometries and topology. The most interesting thermodynamics quantity is the complete free energy \cite{Johnson:2020mwi, Johnson:2021rsh}. In the language of condensed matter/statistical physics, this is sometimes called the ``Quenched'' free energy: $F_{q}(\beta)=-\beta^{-1}\langle\log  Z(\beta)\rangle$ which has a precise distinction with the ``Annealed'' free energy: $F_{a}(\beta)=-\beta^{-1}\log \langle Z(\beta)\rangle$. The distinction is also there in the case of gravitational theories, as the statistical interpretation remains the same for the holographic dual. Usually, free energy is studied at a higher temperature where the high energy configurations dominate the microstates, and there is no robust difference between annealed and quenched free energy \cite{Johnson:2021rsh}. However, in general, one should compute the quenched free energy in order to explore the full thermodynamics properties of that quantum gravity theory.\par
\textcolor{black}{Over the past few years, $T\Bar{T}$, and $J\Bar{T}$ deformations (irrelevant) become a growing interest in high-energy communities. The first study of $T\Bar{T}$ deformation in the context of quantum field theory (in 1+1 dimensions) has been done by Zamolodchikov \cite{Zamolodchikov:2004ce}. Later, it has been extended for other irrelevant deformations in 1+1 dimensions by Smirnov and Zamolodchikov \cite{Smirnov:2016lqw} and the energy spectrum of deformed theory has been computed exactly \cite{Smirnov:2016lqw,Cavaglia:2016oda}. It turns out that the deformations are integrable, i.e. the theory has infinitely many local integrals of motion. $T\Bar{T}$ has found interesting application in holography \cite{McGough:2016lol}. Another interesting irrelevant deformation, namely, $J\Bar{T}$ deformation of 1+1 dimensional CFT, has been explored in \cite{Guica:2017lia}. It is shown to preserve the $SL(2,R)_L\times U(1)_R$ of the original original global conformal group. There are several follow-up works involving the computation of $T\Bar{T}$ and  $J\Bar{T}$ deformed partition function, energy spectrum as well as correlation functions in holographic field theories have been done in \cite{Shyam:2017znq,Kraus:2018xrn,Cottrell:2018skz,Giveon:2017nie,Asrat:2017tzd,Giribet:2017imm, Dubovsky:2018bmo, Datta:2018thy, Aharony:2018bad, Cardy:2019qao, He:2019vzf, He:2019ahx, Mazenc:2019cfg, He:2020udl, Ebert:2020tuy, He:2020qcs, Hirano:2020ppu,Apolo:2019zai,Gorbenko:2018oov}}\footnote{\textcolor{black}{This list is by no means exhaustive. For more details interested readers are referred to the following lecture note \cite{Jiang:2019epa} and two theses \cite{2021arXiv211202596A, Katoch:2023etn} and the references therein. Furthermore, for $T\Bar{T}$ deformation in general dimensional, interested readers are referred to, e.g \cite{Cardy:2018sdv,Hartman:2018tkw, Taylor:2018xcy,Caputa:2019pam,Banerjee:2019ewu}.}}. 
%It turns out that one of the signs of coupling of $T\Bar{T}$ is a bad one and gives a complex spectrum of the deformed theory. 
\par 
The $T\Bar{T}$ is a composite operator made up of the stress tensor and its antiholomorphic counterpart, whereas the $J\Bar{T}$ is a composite operator made up of the stress tensor and the $U(1)$ current operator of the theory. As these operators are composite operators, they should be defined by point-splitting methods in $d>1$. However, in $d=1$, in the absence of spatial direction, they are well defined by construction. The $T\Bar{T}$ deformed (of the dual boundary theory) energy spectrum and the free energy in JT-gravity (specifically for the Airy model) have been studied in \cite{Ebert:2022gyn}. The free energy computation for deformed form factor has been studied in \cite{Alishahiha:2020jko}. \textcolor{black}{Inspired by the recent development of JT-gravity with matter \cite{Jafferis:2022wez}, we take a 2D-gravity non-minimally coupled with $U(1)$ matter field and compute the partition functions along the line of \cite{Iliesiu:2019lfc}. \textcolor{black}{The presence of $U(1)$ matter also enables us to study the effect of $J\Bar{T}$ deformation using holographic duality.} We compute the deformed partition function and its genus expansion \textcolor{black}{by making a $T\Bar{T}+J\Bar{T}$ deformation of dual boundary theory and using the same deformation kernel on the bulk (undeformed) partition function.} \textcolor{black}{Keeping in mind the potential importance of the contributions coming from higher genus topology at a late time following from the works \cite{Saad:2019pqd,Saad:2022kfe,Blommaert:2022lbh}, we also include some higher genus contributions to this deformed partition function (from the gravity side).}  Then, we compute the quenched, annealed free energy and compare it with the undeformed case. After that, we turn our attention to the Spectral Form Factor (SFF) (both the disconnected and connected ones), which utilizes the deformed partition function. We also provide a comparative study of deformed and undeformed ones.} \textcolor{black}{One of the biggest motivations for this analysis comes from the following question: Is there exist a dual RMT description for this deformed JT gravity non-minimally coupled with matter like the pure one \cite{Saad:2019lba}? The answer is unclear and requires much more computations \cite{Rosso:2020wir}. In this paper, we take a step in that direction by analyzing the SFF. %it is still unclear whether the predictions from the $T\bar{T}$ deformed double-scaled RMT will match those coming from the gravity side after including the higher genus contributions to the partition function.
The SFF expects to display a universal ramp-plateau behaviour \cite{Cotler:2016fpe} for theories like JT gravity which admits a dual RMT description signifying a late time chaotic behaviour of the underlying theory. As will be discussed later, for our case SFF also displays similar behaviour perhaps indicating that, there is a dual RMT description, although the precise form of the RMT potential still needs to computed.} Furthermore, our analysis of SFF by incorporating these deformations helps us to test how robust this behaviour is. These deformations have a holographic interpretation in terms of an AdS background with a finite radial cutoff. It makes the analysis of the free energy and SFF  all the more interesting for a bulk 2d near-extremal black hole, keeping in mind that their behaviour can be interpreted as the universality of the spectrum for higher-dimensional AdS black holes with known CFT duals in the large charge and low-temperature regime even when one considers a higher-dimensional black hole\cite{Iliesiu:2019xuh}. \par 
The paper is organized as follows:  In section (\ref{sec2}), we briefly review how to compute the partition function in JT-gravity and the genus expansion of the $n$-point correlation of the partition function. In sections (\ref{section 3}) and (\ref{sec4}), we first discuss how one can arrive at two-dimensional JT-gravity non-minimally coupled with $U(1)$ gauge theory by dimensional reduction of Einstein-Maxwell theory in four dimensions. Then, we move on to the computation of the partition function and its genus expansion in the presence of an abelian charge. In section (\ref{sec5}), we briefly review the $T\Bar{T}$ and $J\Bar{T}$ deformations and its consequences. In section (\ref{sec6}), we compute the deformed partition function using the deformation kernel in a specific limit $h \rightarrow \infty$ with non-zero chemical potential $\mu$. We also provide the derivations of the genus expansion of one and two-point functions, respectively. In section (\ref{sec7}), we start investigating the thermodynamic properties of the deformed theory. We compute the ``Annealed'' and ``Quenched'' free energy of the deformed theory. We make a quantitative comparison between the undeformed and deformed ones. We also investigate these quantities for different deformation parameters. In section (\ref{sec8}), we introduce the notion of Spectral Form Factor (SFF) in JT gravity. We compute both the connected and disconnected SFF for the deformed theory and also make a comparison with the undeformed theory. We analyze the SFF in early-time and late-time separately. We show that in early-time, the connected SFF shows an oscillating dip and ramp structure; in late-time, it shows a plateau in \textit{tau-scaling} limit. We also comment on the main differences between the connected and disconnected SFFs of JT gravity and deformed JT gravity nonminimally coupled with $U(1)$ gauge field. Finally, in section (\ref{sec9}), we summarize our main findings and conclude with some future directions. Some details regarding the computations of the partition function are given in Appendix (\ref{a1}) and (\ref{a2}).

\section{Brief review on JT gravity and genus expansion of partition function}\label{sec2}
The general class of \textit{2D-dilaton gravity} has the following form of the action,
\begin{align}
    S_{2D}=-\underbrace{S_0\chi(\mathcal{M})}_{\text{Topological term}}-\frac{1}{8\pi G_N}\Big[\frac{1}{2}\int_{\mathcal{M}}d^2 x\sqrt{g}(\varphi\mathcal{R}-2U(\varphi))+\int_{\mathcal{\partial M}}dx\sqrt{h}\varphi\mathcal{K}\Big 
]\label{2.1}
\end{align}
which is parametrized by a single function $U(\varphi)$, called the dilaton potential and $\mathcal{K}$ is the Extrinsic curvature. The Jackiw-Titelboim (JT) gravity corresponds to a specific choice of the dilation potential $U(\varphi)=-\Lambda \varphi$ \cite{Jackiw:1984je, Teitelboim:1983ux},
\begin{align}
    \begin{split}
        S_{\textrm{JT}}=-\frac{1}{2}\int_{\mathcal{M}} d^2x \sqrt{g}\varphi (\mathcal{R}-\Lambda) -\int du \sqrt{\gamma}_{uu}\varphi\, \mathcal{K}
    \end{split}
\end{align}
where $\Lambda$ is the cosmological constant of the model. In this paper, we focus on the asymptotically anti-de Sitter ($AdS$) space with cosmological constant $\Lambda=-\frac{2}{L^2}$, with the $AdS$ length $L$. We will set $L=1$. for subsequent analysis. We can now write down the JT action on a manifold $\mathcal{M}$ having boundary $\partial \mathcal{M}$ as,
\begin{align}
    \begin{split}
        S_{\textrm{JT}}[\varphi,g]=-\frac{1}{2}\int_{\mathcal{M}}d^2 x \varphi(\mathcal{R}+2)-\int_{\partial \mathcal{M}}dx \sqrt{h}\varphi (\mathcal{K}-\underbrace{1}_{\text{Counter term}})\,.
    \end{split}
\end{align}
Here we have also added the counter term $-\int_{\partial \mathcal{M}}dx \sqrt{h}\varphi\,$ also.
The $n$-point function can be written in terms of a sum of topologies with genus $g$ and $n$ boundaries,\\ 
  \begin{equation}
      \Big\langle \prod_{i=1}^{n} Z(\beta_i)\Big\rangle_{\text{conn}}\equiv \sum_{g=0}^{\infty}\begin{minipage}
          [h]{0.15\linewidth}
	\vspace{4pt}
	\scalebox{2}{\includegraphics[width=\linewidth]{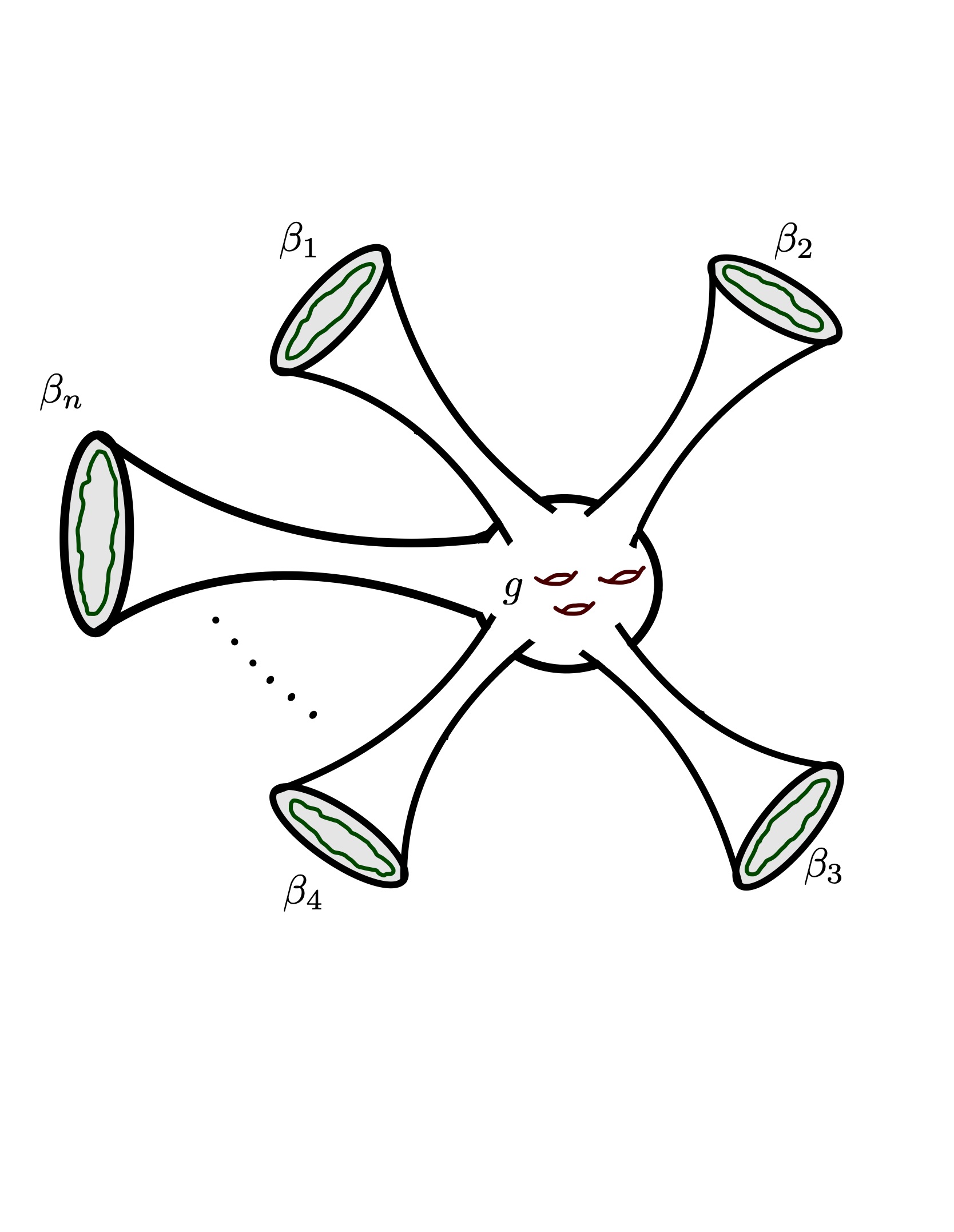}}
      \end{minipage}\hspace{1 cm}\sim \sum_{g=0}^{\infty}\frac{Z_{g,n}(\beta_1,..,\beta_n)}{(e^{S_0})^{2g+n-2}}\,. \label{eq2.4}
  \end{equation}
  The $g$ genus and $n$ boundary  partition function can be evaluated by integrating over the dilaton and the metric,
  \begin{align}
      \begin{split}
          Z_{g,n}(\beta_1,..,\beta_n)=\int \frac{\mathcal{D}\varphi \, \mathcal{D}g_{\mu\nu}}{\text{Vol(diff.)}} \,e^{-S_{JT}[g,\varphi]}\,.
      \end{split}
  \end{align}
  
Here, $\frac{\mathcal{D}g_{\mu\nu}}{\text{Vol.(diff.)}}$ is the diffeomorphism invariant integral measure which takes care of the diffeomorphism redundancy of the theory. The integral over the dilaton field is equivalent to setting $R=-2$ in the action, and we are left with,
\begin{align}
    \begin{split}
         Z_{g,n}(\beta_1,..,\beta_n) =\int_{\text{given topology}(g,n)} \mathcal{D}g_{\mu\nu} \,e^{\int_{\partial \mathcal{M}}\sqrt{h}\varphi(K-1)}\,.
    \end{split}
\end{align}
The integral over metric can be divided into the integral over the wiggly boundaries as shown in the figure in Eq.~(\ref{eq2.4}) and the integral over bulk moduli space.
\begin{align}
    \begin{split}
         Z_{g,n}(\beta_1,..,\beta_n) \sim \int \mathcal{D}(\text{bulk moduli})\int \mathcal{D}(\text{boundary wiggles})\,e^{\int_{\partial \mathcal{M}}\sqrt{h}\varphi(\mathcal{K}-1)}\,.
    \end{split}
\end{align}
The integral over the boundary wiggles will give the disk and trumpet partition function. Using the usual gluing techniques, one can obtain the formula for the partition function for general $g$ and $n$.
\begin{align}
    \begin{split}
     &   Z_{0,1}(\beta)=Z_{\textrm{Sch}}^{\textrm{disk}}(\beta)\,,\\ &
     Z_{g,n}(\beta_1,..,\beta_n)=\alpha^n \int_{0}^{\infty}\prod_{i=1}^{n}b_i db_i\,\mathcal{V}^{\alpha}_{g,n}(b_1,..,b_n)Z_{\textrm{Sch}}^{\textrm{trumpet}}(\beta_i,b_i)\,.
    \end{split}
\end{align}
In the later part of this paper, we will need mostly the one- and two-point functions, expressions of which we write below,
\begin{align}
    \begin{split}
    &    \Big\langle Z(\beta)\Big\rangle=e^{S_0}Z_{\textrm{Sch}}^{\textrm{disk}}+\sum_{g=1}^{\infty}e^{(1-2g)S_0}\int_{0}^{\infty}db\,b\,\mathcal{V}_{g,1}(b) Z_{Sch}^{\textrm{trumpet}}(\beta,b)\,,\\\ &
       \Big \langle Z(\beta_1) Z(\beta_2)\Big\rangle_{\text{conn}}=Z_{0,2}(\beta_1,\beta_2)+\sum_{g=1}^{\infty}e^{-2g S_0}Z_{g,2}(\beta_1,\beta_2)
        \label{2.9}
    \end{split}
\end{align}
with,
\begin{align}
    \begin{split}
        Z_{0,2}(\beta_1,\beta_2)=\int_{0}^{\infty} db\,b\, Z_{Sch}^{\textrm{trumpet}}(\beta_1,b)Z_{Sch}^{\textrm{trumpet}}(\beta_2,b)\,.
    \end{split}
\end{align}
The disk and trumpet partition function in Schwarzian theory \cite{Saad:2019lba} is given by,
\begin{align}
    \begin{split}
        Z_{\textrm{disk}}^{\textrm{Sch}}(\beta)=\frac{e^{\pi^2/\beta}}{4\sqrt{\pi}\beta^{3/2}},\,\,\,Z_{\textrm{Sch}}^{\textrm{trumpet}}(\beta,b)=\frac{e^{-b^2/4\beta}}{2\sqrt{\pi \beta}}.
    \end{split}
\end{align}
The one-point function will contribute to the free energy computation and disconnected SFF computations, and the two-point function is needed to compute the connected SFF.
\section{2D dilaton gravity coupled with $U(1)$ gauge field and genus expansion of grand canonical partition function\label{section 3}}
In this section, we discuss the computation of the partition function for the JT gravity coupled with $U(1)$ gauge field. %and the genus expansion of the partition function. The Euclidean action of JT gravity is given by
As stated in (\ref{2.1}) the first term is the purely topological term coming from the Einstein-Hilbert action and does not contribute to the dynamics of the system. In (\ref{2.1}), $\varphi$ is a dilaton field and $\chi(\mathcal{M})$ is the Euler characteristic of the manifold. 
%Rather, in our context, we are interested in JT gravity coupled with a U(1) gauge field.
To construct a consistent coupling with $U(1)$ gauge field, we start from the Einstein-Maxwell theory in asymptotically $AdS_4$ spacetime. The action is given by,
\begin{align}
\begin{split}
 \hat{S}_{EM}=&-\frac{1}{16\pi G_N}\Big[\int_{\mathcal{M}_4}d^4 x\,\sqrt{g(\mathcal{M}_4)} \,(\mathcal{R}+2\Lambda)-2\int_{\partial \mathcal{M}_4}\,d^3u \sqrt{h_{\partial\mathcal{M}_4}}\mathcal{K}\Big]\\ &
  -\frac{1}{4q}\int_{\mathcal{M}_4}d^4 x\,\sqrt{g(\mathcal{M}_4)} F^{\mu\nu}F_{\mu\nu}\,.
    \end{split}\label{3.2}
\end{align}
The variational problem in (\ref{3.2}) is well defined if we fix the components of $A_i$ along the boundary $\partial\mathcal{M}_4$. Such a boundary condition implies that we study the system with a fixed chemical potential. Moreover, we can dimensionally reduce the action from four to two dimensions in a self-consistent manner. We should remember that the dimensional reduction happens only for the low temperature $\beta\ge 1$ where the semiclassical approximation does not work. Introducing a dilaton field $\varphi$ we can dimensionally reduce the action mentioned in (\ref{3.2}) to \cite{Iliesiu:2020qvm},
\begin{align}
    \begin{split}
        & \mathcal{S}_{\textrm{EM}}^{2D}=-\frac{1}{4G_N}\Big[\int_{\mathcal{M}_2}\,\sqrt{g}\Big(\varphi \mathcal{R}-2 U(\varphi)\Big)+2\int_{\partial\mathcal{M}_4}\sqrt{h}\varphi \mathcal{K}\Big]-\frac{1}{4q^2r_0}\int_{\mathcal{M}_2}\,\varphi^{3/2}\mathcal{F}_{\mu\nu}\mathcal{F}^{\mu\nu}+\mathcal{S}[SO(3)]\,.
    \end{split}
\end{align}
The SO(3) gauge field, denoted by $H_{\mu\nu}$, is associated with the diffeomorphisms of the transverse sphere.
Therefore, the partition function is given by
\begin{align}
    \begin{split}
        & \mathcal{Z}(\beta)=\int \mathcal{D}g_{\mu\nu}\mathcal{D}A_\mu \mathcal{D}H_{\mu\nu}\mathcal{D}\varphi \exp{\Big[-S_{JT}+\frac{1}{2}\int_{\mathcal{M}}d^2 x\,\sqrt{g} \,\varphi^{3/2}\mathcal{F}_{\mu\nu}\mathcal{F}^{\mu\nu}+\mathcal{S}[H_{\mu\nu}]\Big]}\,.\label{3.3}
    \end{split}
\end{align}
As we are in two dimensions, the path integral over the gauge field in (\ref{3.3}) can be exactly carried out using the fact that in two dimensions, the partition function of the Maxwell theory or, more generically, the Yang-Mills theory depends only on the area of the manifold and is given by \cite{Iliesiu:2020qvm},
\begin{align}
    \begin{split}
        & \mathcal{Z}(\beta,\mu)=\sum_{Q,j} (2j+1)\,{\chi_j(\mu_{YM})}e^{\beta\mu Q}\int \mathcal{D}g_{\mu\nu}\mathcal{D}\varphi \,e^{-\hat{S}_{Q}[g_{\mu\nu},\varphi]}
    \end{split}\label{3.5}
\end{align}
where,
\begin{align}
    \begin{split}
        \hat{S}_Q=-\frac{1}{4G_N}\Big[\int_{\mathcal{M}_2}\,\sqrt{g}\Big(\varphi\mathcal{R}-2 U_{Q,j}(\varphi)\Big)+2\int_{\partial\mathcal{M}_2}\sqrt{h}\,\varphi\, \mathcal{K}\Big]\,\,\,\,\, 
    \end{split}
\end{align}

$\chi_j(\mu_{YM})$ \text{is the character of SO(3)}
with,\footnote{We ignore the third term in the limit $\frac{r_0}{L}<<1$.}
\begin{align}
    \begin{split}
        U_{Q,j}=r_0\Big[\frac{Q^2}{4\pi\varphi^{3/2}}+\frac{3}{\varphi^{5/2}}j(j+1)\underbrace{-\frac{3}{L^2}\sqrt{\varphi}-\frac{1}{\sqrt{\varphi}}}_{\text{ Coming from dimensional reduction ($\tilde{U}(\varphi_{b,Q})$)}}\Big]\,.\label{3.6m}
    \end{split}
\end{align}
\textcolor{black}{Where Q is the U(1) charge and $j(j+1)$ is the eigenvalue of the SO(3) casimir $J^2$.} Following \cite{Iliesiu:2020qvm}, one can show that, in the near horizon limit expanding the dilaton field around a background value $\varphi_0(Q)$ and introducing fluctuations $\varphi$ , the action (\ref{3.6m}) can be reduced exactly to the  JT action. Now we have all the ingredients to calculate the different correlators involving partition function. For a general Yang-Mills theory coupled with JT gravity, the genus expansion of the point correlation function can be computed using the usual gluing techniques.

As an essential part of gluing techniques, we discuss the  Weil-Peterson volume here.
The Weil-Peterson volume $\mathcal{V}_{g,n}(b_1, \cdots ,b_n)$ of a Reimannn surface $\Sigma_{g,n}$ (i.e with $g$-genus and $n$ distinct marked points $p_j$) defined as \cite{mirzakhanirecursion, Dijkgraaf:2018vnm, Eynard:2007fi}\,,
\begin{equation}
\mathcal{V}_{g,n}(b_1, \cdots ,b_n)=\frac{1}{(2\pi^2)^{(3g-3+n)}}\int_{\bar{\mathcal{M}}_{g,n}} \exp(\omega +\frac{1}{2}\sum_{j=1}^n\psi_jb_j^2)
\end{equation}

\textcolor{black}{where $\mathcal{M}_{g,n}$ is the moduli space of the Riemann surface $\sum_{g,n}$ of \textcolor{black}{ dimension }$(3g-3+n)\,.$ $\bar{\mathcal{M}}_{g,n}$ is the \textit{Deligne-Mumford compactification} of the aforesaid moduli space. $ \psi_j \equiv c_1(\mathcal{L}_j)$ is the first Chern class of the tautological bundle $\mathcal{L}_j$
over the Deligne-Mumford compactification of the moduli space whose fiber at the point $(C,x_1\cdots x_n)\in \bar{\mathcal{M}}_{g,n} $ is the cotangent bundle of curve C at $x_i$. $\omega$ is the symplectic form on $\bar{\mathcal{M}}_{g,n}.$ $b_j$ is the length of the $j$-th geodesic boundary. For Deligne-Mumford compactification $\bar{\mathcal{M}}_{g,n}$ of the moduli space $\mathcal{M}_{g,n}$ can be constructed by adjoining hyperbolic surfaces with simple closed geodesics of length close to zero.
\textcolor{black}{For any set $(d_1,\cdots ,d_n)$ of non-negative integers we define the top intersection number of $\psi$ classes as }\cite{Dijkgraaf:2018vnm} :}
\vspace{-0.4cm}
\begin{align}
\langle \tau_{d_1}\cdots \tau_{d_n}\rangle  =\int_{\bar{\mathcal{M}}_{g,n}} {\prod}_{i=1}^n\psi^{d_i}_i\,.
\end{align}

Now, to quantize our theory, which has an action (\ref{3.3}), we introduce two Lagrange multipliers $\phi_{b}^{U(1)}$ and $\phi_{b}^{SO(3)}$. The action in terms of those can be written as,
\begin{align}
\begin{split}
S_{\textrm{EM}}&=-\frac{1}{4G_N}\Bigg[ \int_{\mathcal{M}} d^2x \sqrt{g}[\varphi \mathcal{R}- 2U(\varphi) ]+2\int_{\partial\mathcal{M}} du \sqrt{h}\varphi ({K}-1) \Bigg]\\&-i\int_{\mathcal{M}} (\phi_{b}^{U(1)}f+ tr(\phi_{b}^{SO(3)}H))
-\int_{\mathcal{M}} d^2x\sqrt{g}\Bigg[\frac{3G_Nr_0}{2\varphi^{5/2}}tr(\phi_{b}^{{SO(3)}^2})+\frac{e^2r_0}{2\varphi^{3/2}}tr(\phi_{b}^{{U(1)}^2})\Bigg]\,.\label{3.9m}
\end{split}
\end{align}

{Here the source current $j(x)$(the coefficient of $\textrm{tr}(\phi_{b}^{U(1)^2})/\textrm{tr}(\phi_{b}^{SO(3)^2})$ \cite{Iliesiu:2019xuh}
can be absorbed by changing the surface term $d^2x
\sqrt{g}$. As the theory is
invariant under diffeomorphisms (local area-preserving), essentially the partition function only depends on the dimensionless quantity \begin{align}\hat{a} =\int d^2x \sqrt{g} j(x)\,.\label{3.10}\end{align}

Therefore, the partition function for a two-dimensional manifold with $g$ handles and $n$ boundaries coupled to a $U(1)\times SO(3)$ theory is given by the path integral over the dilaton field and metric tensor as follows,

\begin{align}
    \begin{split}
        Z_{\textrm{2DEMSO(3)}}^{g,n}&=\int \mathcal{D}g_{\mu\nu}\mathcal{D}\varphi\, e^{-\hat{S}_{Q,\Tilde{U}}}\underbrace{\sum_R (\dim R)^{\chi_{\mathcal{M}_{g,n}}} \chi_R(h_b)\cdots \chi_R(h_n)\sum_{R'}(\dim R')^{\chi_{\mathcal{M}_{g,n}}} \chi_{R'}(h_b)\cdots \chi_{R'}(h_n)}_{\text{Topological factor}}\\&\times\exp\Big({{\frac{C_2(R)}{N_1}}\int d^2x\frac{\sqrt{g}}{2}e^2r_0\varphi^\frac{-3}{2}}\Big)\exp\Big({{\frac{C_2(R')}{N_2}}\int d^2x\frac{3\sqrt{g}}{2}G_Nr_0\varphi^\frac{-5}{2}}\Big)\,.\label{3.12m}
    \end{split}
\end{align}
In (\ref{3.12m}), $C_2(R)$ and $C_2(R')$ represent the quadratic Casimir of the representation for U(1) and SO(3), respectively, and $\chi$ is the character of the irreducible representation, and $h$ is the holonomy defined over the boundary $\partial M$.

\begin{align}h\equiv \mathcal{P}\exp(\oint\mathcal{B}_aT^a)\,.\end{align}
$R,R'$ denote the sum over the representations $U(1)$ and $SO(3)$ respectively. Here $N_1(\equiv 1)$ and $N_2(\equiv \frac{1}{2})$ are the Dynkin indices of U (1) and SO (3), respectively.

Now, to calculate the different thermodynamic quantities such as free energy and SFF, one needs to know how $r_{0}, \varphi_{0}$ depends on $Q.$ For that, we choose the near-extremal limit \cite{Nayak:2018qej, Iliesiu:2020qvm,Bhattacharjee:2020nul, Banerjee:2021vjy}, which solely fixes their functional form. The limit has a well-motivated physical interpretation \cite{Strominger:1996sh}. At $T=0$, we have an extremality condition, and the black hole mass is given by $M_0(T=0)$. We want to see how the thermodynamic quantities change if we slightly switch on the temperature. It can be shown that $E-M_0(T=0)\sim \frac{3}{2}T>T$ for a small non-zero temperature. Now, the important point is that when we calculate the density of state for such a near-extremal black hole, it does not show a finite energy gap if we take the counter-term to be vanishing. \textcolor{black}{But in the case of a $T\bar{T}+J\bar{T}$ deformation, which we will discuss later in the paper, there is an energy gap in comparison to the undeformed theory, which supports the holographic description of switching the $J\bar{T}$ deformation on the AdS background with a finite radial cutoff \cite{Iliesiu:2020zld}. }

%\newpage
\textbf{Near-extremal limit:} We know that charge $(Q)$ determines the width of the throat. The classical solution corresponds to $U(\varphi_{0})=0$, where $U(\varphi_0(Q))$ is defined in (\ref{3.12}). Then, in this near-extremal limit, we get \cite{Iliesiu:2020qvm,Mertens:2022irh}
\begin{align}              
Q^2 =\frac{\pi}{G_N}(r_0^2+3\frac{r_0^4}{L^2})\,,\,\,\,\,\,\,\,\,\,\,\,\,\,\,M_0=\frac{r_0}{4G_N}(1+2\frac{r_0^2}{L^2})\,,\,\,\,\,\,\,\,\,\,\,\,\, S_0=\frac{\pi r_0^2}{4G_N}\,.
\end{align}
 Furthermore, we work in the limit $r_0<<L^2,$ that gives 
 \begin{align}
 r_{0}=\frac{Q}{\sqrt{4\pi}}\,, \quad \varphi_0(Q)=\frac{Q^2}{4\pi}\,, \textrm{along with},\, U'(\varphi_0)=-\frac{4\pi}{Q^2}\label{3.14a}
 \end{align}
 which sets the Ricci scalar $\mathcal{R}\approx -\frac{8\pi}{Q^2}<0$ .\footnote{For more details see Appendix~\ref{a1}.}
 
% Now, at each point in AdS, there is a $S^2$, which is basically the size of the sphere. JT gravity solution arises by setting the radius of the sphere to be large and allowing a small fluctuation around it. 
The near-extremal dynamics is dominated by the modes inside the throat in the near-horizon  (NHR) region. Fluctuations outside the throat are too small to be taken into account. So, we evaluate the far-region (FAR) action on-shell.\footnote{Now considering second-order fluctuations in dilaton field $\varphi$, \textcolor{black}{\textit{ we integrate over all possible hyperbolic surfaces with $R=-\frac{2}{L_2^2}$}. One should be careful about throwing away terms due to small fluctuations in the FAR regime because there is a smooth matching at the boundary of the NHR and FAR regime.}} In principle, one should have

$$g_{\mu\nu}=g_{\mu\nu}^{\textrm{ext}}+\delta g_{\mu\nu}^{\textrm{near ext}}\,, \,\,\,\,\,\,\, \,\,\,\,\,\,\, \,\,\,\,\,\,\,\,\,\,\,\, \,\,\,\,\,\,\, \varphi=\varphi^{ext}+\delta\varphi^{\text{near ext.}}\,.$$
The expansion in the FAR regime around the extremal solution gives rise to a boundary term from the action \cite{Iliesiu:2020qvm}\,,

\begin{align}
S^{Q,j}_{\textrm{FAR}}[g_{\mu\nu},\varphi]=S^{Q,j}_{\textrm{FAR}}[g_{\mu\nu}^{ext.},\varphi^{\textrm{ext}}]-\frac{1}{2G_N}\int_{\partial M_{\textrm{NHR}}}du\sqrt{h}[\varphi \delta \mathcal{K}-(\partial_n\varphi-\varphi \mathcal{K})\delta\sqrt{h_{uu}}]
\end{align}
where $$\delta \mathcal{K}\equiv \mathcal{K}_{\textrm{NHR}}-\mathcal{K}_{\textrm{ext.}}\,.$$
Also we have imposed Dirichlet boundary conditions $\delta\sqrt{h_{uu}}=0$.\par
So, in general, the dynamics of the near-extremal black hole come from the following action,
\begin{align}
\begin{split}
S_{\textcolor{black}{\textrm{2DEM}}}^{Q,j}&= \beta M_0(Q,j)-\frac{1}{4}
\int_{M_{\textrm{NHR}}}d^2x\sqrt{g}\Bigg[\varphi_0\mathcal{R}+\varphi(R+\frac{2}{L_2^2})+\mathcal{O}\Big({\frac{\varphi^2}{\varphi_0^2}}\Big)\Bigg]\\&-\frac{1}{2}\int_{\partial M_{\textrm{NHR}}}du \sqrt{h}\Bigg[\varphi_0(Q,j)\mathcal{K}_{\textrm{NHR}}+\frac{\varphi_{b,Q}}{\epsilon}(\mathcal{K}_{\textrm{NHR}}-\frac{1}{L_2})\Bigg]\,.
\end{split}
\end{align}
Here $M_0(Q,j)$ is the extremal mass as defined in \cite{Iliesiu:2020qvm}. The extrinsic curvature arises only from the FAR regime. \textcolor{black}{For further discussion regarding the Near-Horizon regime (NHR) and Far-Horizon regime (FAR)}, we refer the reader to \cite{Iliesiu:2020qvm}. However, we now have a complicated dilaton potential $U(\varphi)$. \textcolor{black}{Now, we compute the path integral over hyperbolic surfaces with $R=-\frac{2}{L_2^2}$, with asymptotically $AdS_2$ boundary, keeping terms up to quadratic in dilaton.} Details of the computation are given in Appendix~\ref{a1}. 
%However, one should remember that the saddle point condition eventually fixes the Ricci scalar that could be inserted as a delta function constraint in the gravitational path integral, as discussed in the .

While calculating the partition function, we encounter UV divergences. Therefore, we add the defect action (counterterm action) arbitrarily close to the boundary to cancel those divergences such that the path integral converges.\footnote{We discuss it further later in section \ref{sec4}\,.}. The defect action is given by\footnote{Here, $\epsilon \rightarrow 0$ and $p\rightarrow \infty\,.$},

%Now we get a term from $\frac{P(\varphi_0(Q))}{L(\varphi_0(Q))}$ in (\ref{3.13}) proportional to \textcolor{black}{simplify this expression. $r_0$ will cancel etc like in (3.15)}, $$\sim \frac{-\frac{C_2 e^2 r_0 \varphi _0^{-3/2}}{N}+\frac{\left(6 r_0\right) \sqrt{\varphi _0}}{L^2}+2 r_0 \frac{1}{\sqrt{\varphi _0}}}{-\frac{(3 C_{2}) e^2 r_0 \varphi _0^{-5/2}}{2 N}-\frac{\left(3 r_0\right) \frac{1}{\sqrt{\varphi _0}}}{L^2}+r_0 \varphi _0^{-3/2}}\,.$$ 
%Therefore, the defect (counterterm) action one needs to add at the near boundary is given by \cite{Iliesiu:2020qvm}\,

\begin{align}S_{\text{defect}}=2\int_{I}du \sqrt{g_{uu}}\,\Bigg[\epsilon \Big({c} Q^2\Big)-{\varphi_0(Q)} \Bigg]\textrm{tr}(\phi_{b}^{(U(1))})^2 \,.\label{eq 3.13} \end{align}
\textcolor{black}{ The defect action in (\ref{eq 3.13}) is the most complete one to take care of all the divergence while considering the second-order fluctuations in the dilaton field. While integrating over the dilation field (\ref{A.4}), we get a divergence term coming from the Gaussian functional integral, which is obvious in the continuum limit. We should remember that the information on the second-order fluctuations is encoded into $U''(\varphi_0)$ and this is the source of this divergence inside the path integral. Therefore, we use the \textit{Zeta-function regularization} to extract the convergent part of the infinite product.}
In (\ref{eq 3.13}), $I$ is arbitrarily close to the boundary, and {${c}$ is an arbitrary constant $\ge 0$ to make the sum over $Q$ in the path-integral convergent. We define the following, $$K_0(Q)\equiv c\, Q^2\,, \textcolor{black}{G_N\equiv \frac{1}{4}}\,, \quad \textrm{and} \quad \Tilde{M}_0(Q)\equiv M_0(Q)+c\, Q^2\,. $$  We proceed by choosing the \textit{Dynkin index,} $N_1 \equiv 1$
. Finally the one-point function can be written as

\begin{align}
\begin{split}
   \langle {Z}(\beta,\mu)\rangle = e^{S_0}
  \begin{minipage}[h]{0.07\linewidth}
	\vspace{4pt}
	\scalebox{1.2}{\includegraphics[width=\linewidth]{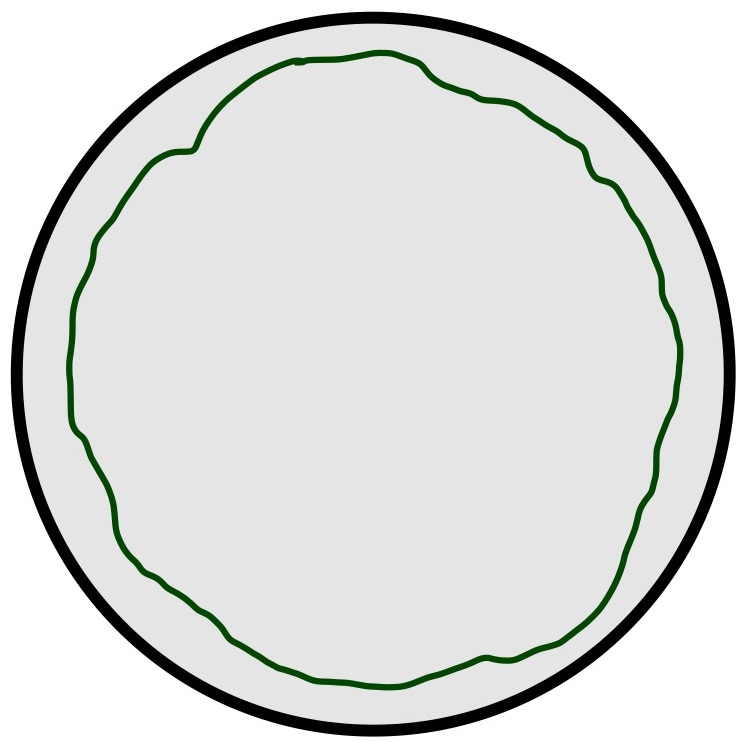}}
   \end{minipage}\hspace{0.3 cm}
  + \int d\Tilde{h}\sum_{g=1}^{\infty}e^{(1-2g)S_0}\int_{0}^{\infty}bdb \,\, \begin{minipage}[h]{0.12\linewidth}
	\vspace{4pt}
	\scalebox{1.1}{\includegraphics[width=\linewidth]{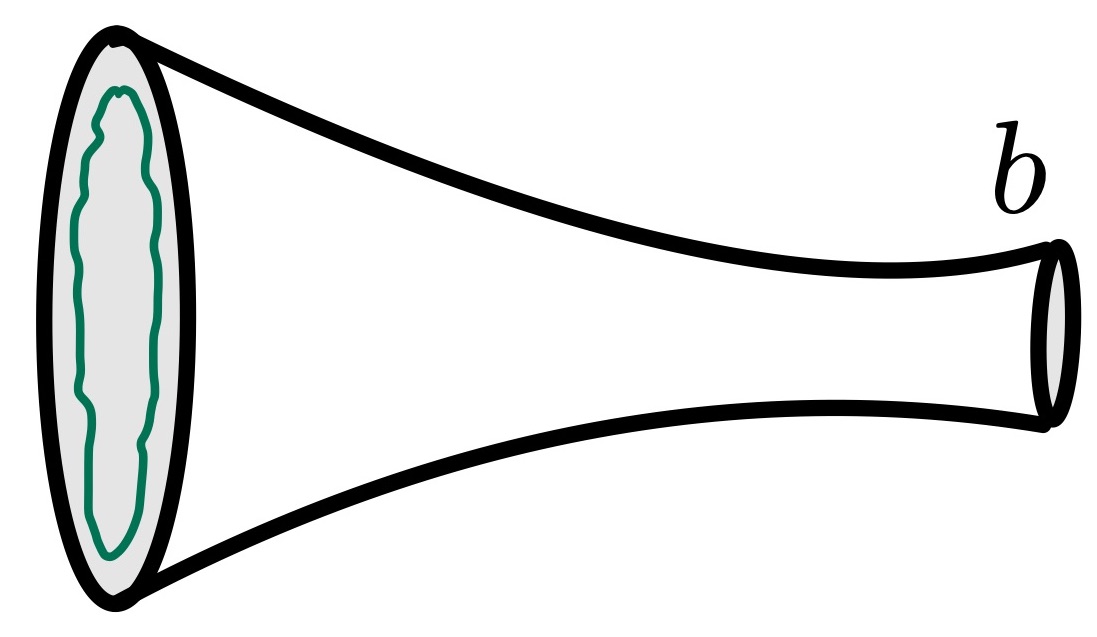}}
   \end{minipage} \,\hspace{0.1 cm}
    \begin{minipage}[h]{0.07\linewidth}
	\vspace{15 pt}
	\scalebox{2.2}{\includegraphics[width=\linewidth]{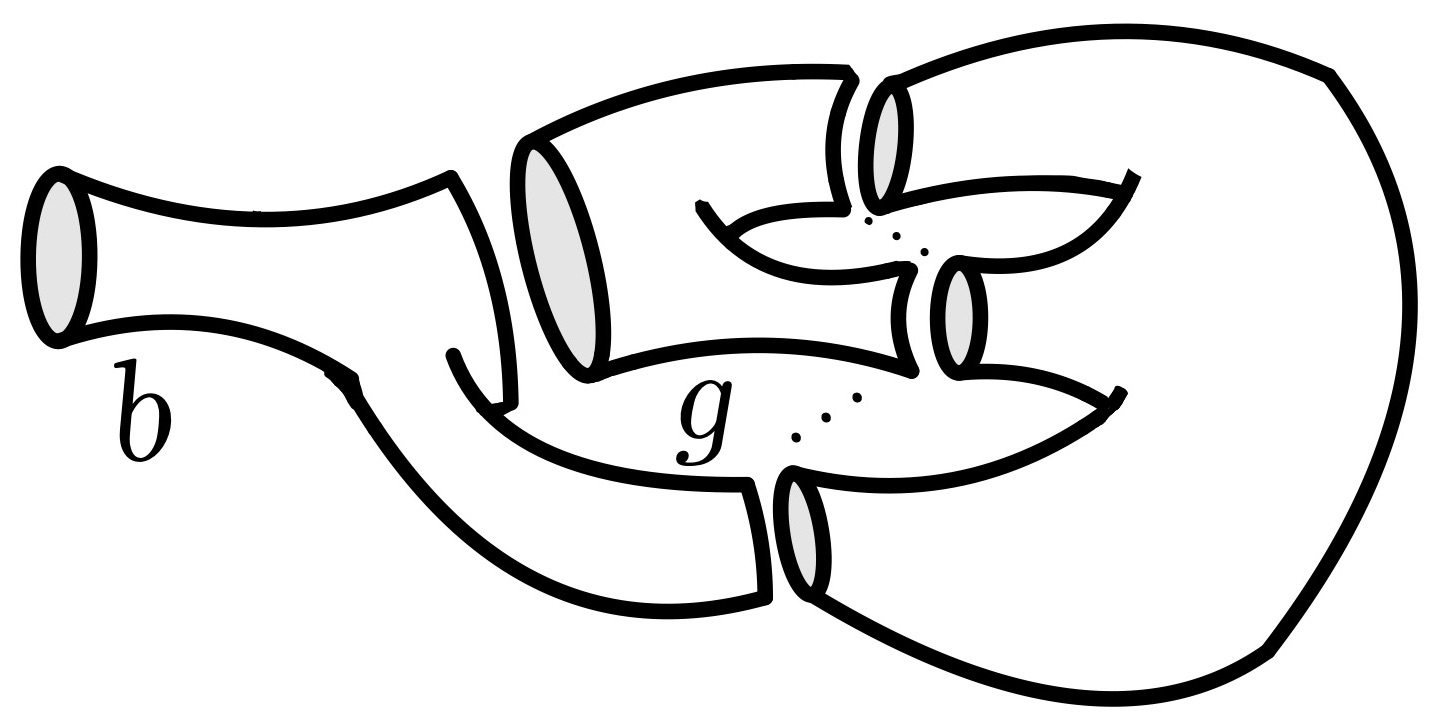}}
   \end{minipage} \hspace{1 cm}
   \label{3.16}
   \end{split}
\end{align}

The first term in (\ref{3.16}) refers to the disk partition function, and the next term denotes the gluing of the trumpet to the moduli space of hyperbolic metric.

\textcolor{black}{Therefore, by doing the path integral in (\ref{3.12m})\footnote{The path integral has been done using the action (\ref{3.9m})\,. } and using the result obtained in (\ref{A1}) and (\ref{A2}), we can write the partition function for $n=1$ {bordered} Riemann surface, with $g$-genus in the near-extremal limit as,}

\vspace{0.3 cm}

\begin{align}
\begin{split}
      Z_{\textrm{2DEMSO(3)}}^{g,n=1}(b,\tilde h)\equiv& \begin{minipage}[h]{0.07\linewidth}
	\vspace{12 pt}
	\scalebox{2.5}{\includegraphics[width=\linewidth]{JTEM.jpg}}
   \end{minipage}\hspace{1.5cm}= \underbrace{\sum_R (\dim R)^{\chi_{\mathcal{M}_{(g,1)}} }\chi_R(h_b)\sum_{R'}(\dim R')^{\chi_{\mathcal{M}_{(g,1)}} } \chi_{R'}(h_b)}_{Topological factor}\,,  \\& \color{black}\,\hspace{3 cm}\times {\mathcal{V}_{g,1}(b)}\Bigg(\frac{U''(Q)}{\pi}\Bigg)^{1/4}\exp\Bigg\{4\pi\chi \varphi_0(Q)\Bigg\}\,,\\ &
  = \sum_{Q} \sum_{j}(2j+1)^{\chi_{\mathcal{M}_{(g,1)}} } \chi_{j}(h_b)\,
   \color{black}\,\times {\mathcal{V}_{g,1}(b)}\Bigg(\frac{U''(Q)}{\pi}\Bigg)^{1/4}\exp\Bigg\{\underbrace{4\pi\chi \varphi_0(Q)}_{\chi{S}_0(Q)}\Bigg\} \nonumber 
   \end{split}
   \end{align}
   \begin{align}
   \begin{split}
  \hspace{3 cm} &  \\&
=\sum_{Q}\sum_{j}(2j+1)^{\chi_{\mathcal{M}_{(g,1)}} } \chi_{j}(h_b)  \,\times {\mathcal{V}_{g,1}(b)}\Bigg(\frac{U''(Q)}{\pi}\Bigg)^{1/4} e^{\chi_{\mathcal{M}_{g,n}}{S}_0(Q)}\\&
   \end{split} \label{3.13}
\end{align}
where $\varphi_0(Q)$ is defined in \eqref{3.14a} and,
\begin{align}   
U(\varphi)\equiv \,\,2\alpha'\varphi^{\frac{-3}{2}}+\underbrace{\tilde{\beta}}_{\text{0 in j=0 sector}}\varphi^{\frac{-5}{2}}+\gamma\varphi^{\frac{1}{2}}+\delta\varphi^{\frac{-1}{2}}\,\,\,\,\,\,\,\, \textrm{with} \,\,\,\,\,U''(\varphi_0(Q))>0\label{3.12}
\end{align}

with the parameters defined as follows, $$ \tilde{\beta}=\frac{C_2(R')}{N}\frac{3}{2}G_Nr_0(Q),
\,\,\delta= -2r_0(Q),\,\, \gamma=-\frac{6r_0(Q)}{L^2},\,\, \alpha'=\frac{C_2(R)}{2N}Q^2r_0(Q),\,\, $$
and,
$$ -U'(\varphi_0(Q))= 3\alpha'\varphi_0(Q)^{-5/2}  +\frac{5}{2} \tilde{\beta  }\varphi_0(Q)^{-7/2} -\frac{\gamma}{2} \varphi_0(Q)^{-1/2} +\frac{1}{2}  \delta  \varphi_0(Q)^{-3/2}.$$
Here $r_0$ is the black hole horizon radii and $L$ is the usual AdS radius. \textcolor{black}{Now we proceed to calculate the trumpet and the disk partition function in near-extremal limit in the following section. }

\section{The Disk and Trumpet geometry in $Schw.\times U(1)\times SO(3) \text{ with } (j_{SO(3)}=0)$}\label{sec4}

In recent times, it has been shown that the, dual boundary theory of $\textrm{JT}$ gravity on an asymptotically AdS spacetime can be thought of as a $0+1$ dimensional Schwarzian theory which is the low energy effective action of SYK model in large $N$ limit \cite{1993PhRvL..70.3339S, Kitaev:2017awl}. The SYK model is a solvable quantum mechanical system with a well-behaved large $N$ limit (it classicalizes in this limit) and maximally chaotic  (it was shown to saturate the chaos bound proposed in \cite{Maldacena:2015waa}). In the infrared regime, this model develops an approximate conformal symmetry. Because of this, in recent times, it has played an important role in understanding several aspects of holography in context of $AdS_2$ holography \cite{Maldacena:2016hyu}. 

The action of the dual one-dimensional boundary theory \cite{Blommaert:2018oro} is given by.
\begin{align} \label{schwarz}
S[f]&=-C\int_0^\beta du \Bigg( \{f,u\}+\frac{2\pi^2}{\beta^2}f'^2 \Bigg)\,,\\&
=-C\int_0^\beta du\{F,u\}\,,
\end{align}

 where $C$ is the coupling constant of the {zero-temperature} theory and \textcolor{black}{$F(u)=\tan(\frac{\pi f(u)}{\beta})$}. Here $f(u+ \beta) = f(u ) + \beta$ runs over the space $\textrm{Diff}(S^1)$
of diffeomorphisms on the thermal circle around the boundary, and $\{f,u\}$ denotes the Schwarzian. 
\begin{align}
\{f,u\}=\frac{f'''}{f'}-\frac{3}{2}\frac{f''}{f'^2}\,.
\end{align}

Now, we focus on the disk partition function. Following \cite{Blommaert:2018oro}, after performing the path-integral over the gauge field, we obtain the following, \footnote{ Interested readers are referred to \cite{2dg, Blommaert:2020hgi} for more details. }
\begin{align}
\Tilde{Z}_{\textrm{disk}}(g,h)&=\sum_{R,a,b} \textrm{dim} (R)R_{ab}(g)R_{ba}(h^{-1})e^{-T H(R)}\,,\\&
=\sum_R \textrm{dim}(R)\chi_R(U)e^{-T H(R)}\,.
\end{align}
Here, $\chi_R(U)=\textrm{Tr}R(U)$ is the character of the representation, and $U=g\cdot h^{-1}$ is the holonomy along the entire boundary of the disk. $|g\rangle$ and $|h \rangle$ are the boundary states. 
%which can be written as a sum over states on a representation basis. 
Essentially, for topological BF theory the
boundary-propagating Hamiltonian factor calculated in a state $|R, ab\rangle$ is given by,
\begin{equation}
    TH(R)=L\frac{\mathcal{C}(R)}{C}\,.
\end{equation}
with normalized wavefunctions defined as,
$$\langle g| R, ab\rangle =\sqrt{\textrm{dim} R} \, R_{ab}(g)$$
where $R_{ab}(g)=\langle a | g| b \rangle$ is the matrix element in  representation R.

Then using (\ref{3.10}) we get,  \begin{align}
\begin{split}
\Tilde{Z}_{\textrm{disk}}(\hat{a},g,h)&= \sum_R \textrm{dim}(R) \, \chi_R(U)e^{-\hat{a}\,\mathcal{C}(R)}\,,\\&
= \sum_R \textrm{dim}(R) \, \chi_R(U)e^{-\mathcal{C}(R)\int_{\mathcal{M}} d^2x \sqrt{g} j(x)}\,. \label{neweq}
\end{split}
\end{align}

The path integral over the metric depends on the boundary degrees of freedom of $AdS_2$. We use the Poincare coordinates,
$$ ds^2= \frac{dF^2+dz^2}{z^2}, \,\,\,\,\,\,\,\,\,   \,\,\,\,\,\,\,\,\, g_{uu}|_{bdy.}=\frac{F'^2+z'^2}{z^2}=\frac{1}{\epsilon^2} \,.$$
Here, the boundary is parametrized using the variable $u$, with $F'=\frac{\partial F}{\partial u}$. Solving the second equation mentioned above, we get $z=\epsilon F'+\mathcal{O}(\epsilon^2)$. Since $z(u)$ is small in $\epsilon \rightarrow 0$, the path integral has a major contribution from the patches near the boundary. Furthermore, the extrinsic curvature takes the following form, \begin{align}
\mathcal{K}[F(u),z(u)]=\frac{F'(F'^2+z'^2+zz'')-zz'F''}{(F'^2+z'^2)^{3/2}}=1+\epsilon^2\,\{F,u\}+\mathcal{O}(\epsilon^3).
\end{align}

Using the fact that $\mathcal{R}=2U'(\varphi_0(Q)))$ is a constant, we get the following, 

\begin{align}
\tilde{C} \int_{\mathcal{M}}d^2x\sqrt{g}\,\,\,=\,\,\, \frac{\tilde{C}\int_{\mathcal{M}}d^2x \sqrt{g}\mathcal{R}}{2U'(\varphi_0(Q))}=\,\,\,\frac{-\tilde{C}}{U'(\varphi_0(Q))}\Bigg[\int_{\partial \mathcal{M}}\sqrt{h}\mathcal{K}-2\pi \chi(\mathcal{M})\Bigg]\,.\label{4.9}
\end{align}
For the disk, the Euler characteristic is $\chi(\mathcal{M})=1$ and we have used the Gauss-Bonnet relation, 
 \begin{align}
 \frac{1}{2}\int_{\mathcal{M}}\sqrt{g}\mathcal{R}+\int_{\partial \mathcal{M}}\mathcal{K}=2\pi \chi(\mathcal{M})\,.
 \end{align}\\
In(\ref{4.9}) $\Tilde{C}$ is a non-zero arbitrary constant.
%\vspace{0.3cm}

 \textbf{ The disk partition function:} The Euler-characteristic for the disk is $\chi_{\textrm{disk}}=1$. Now given (\ref{neweq}}), to compute the disk partition function, we now have to perform the integral over the metric and dilaton field. To proceed further, we write down the metric for the disk in the following way: \begin{equation}
   ds^2=d\rho^2+\sinh^2(\rho)d f^2\,.
 \end{equation} 
 $f(u)$ describes the wiggly boundary, where $u$ is a rescaled proper length coordinate along the boundary, ranging from zero to $\beta$. $\rho$ denotes the coordinate such that the induced metric on the boundary becomes $g_{uu}|_{bdy}=\frac{1}{\epsilon^2}$. \footnote{The boundary value of the dilaton field becomes $\varphi|_b=\frac{\varphi_{b,Q}}{\epsilon}$, Where from near-extremal limit $\varphi_{b,Q}\sim Q^3$,which we used in the later sections during plotting.} Now, the following expression gives the disk partition function, 

 \begin{align}
 \begin{split}
Z_{\textrm{2DEMSO(3)}}^{\textrm{disk}}&=\int  \mathcal{D}\,\varphi\mathcal{D}g_{\mu\nu} e^{-\hat{S}_{Q,\tilde{U}}}\underbrace{\sum_{R'}\textrm{dim}(R') \chi_{R'}(h_{AdS_2})\sum_R \textrm{dim} (R)\,\chi_R(h_{AdS_2})}_{\text{Topological factor}} \underbrace{e^{-\beta \Tilde{M_0}(Q)}}_{\text{mass term}}\\ & \hspace{1 cm}\times\exp\Big(\textstyle{{{\frac{C_2(R)}{N}}\int d^2x\frac{\sqrt{g}}{2}q^2r_0\varphi^\frac{-3}{2}}}\Big)\,,\\&
=\underbrace{\int\frac{\mathcal{D}\mu(F)}{\textrm{Vol({SL(2,R))}}} \exp\Big(-{\frac{1}{2G_N}\int_0^\beta\frac{{\varphi_{b,Q}}}{{\epsilon^2}}(1+\epsilon^2 \{F,u\}\,du})\Big)}_{\int ds \frac{s}{2\pi^2}\sinh(2\pi s)\exp(\frac{-\beta s^2}{2\varphi_{b,Q}})=\frac{\varphi_{b,Q}^{3/2}}{\sqrt{2\pi\beta^3}}\exp(\frac{2\pi^2}{\beta}\varphi_{b,Q})}\\& \hspace{0.5 cm}
\times\int \mathcal{D}\varphi  \exp\Big({{\frac{C_2(R)}{N}}\int d^2x\frac{\sqrt{g}}{2}q^2r_0\varphi^\frac{-3}{2}}\Big)e^{-\hat{S}_{Q,\Tilde{U}}}{e^{-\beta \Tilde{M_0}(Q)}}\times(\text{Topological factor})
\end{split}
 \end{align}

%\vspace{-0.5 cm}
yielding,

\begin{align}
\begin{split}
{Z}_{\text{2DEMSO(3)}}^{\text{disk}}\equiv\hspace{- 0.1 cm}\begin{minipage}[h]{0.15\linewidth}
	\vspace{4pt}
	\scalebox{0.7}{\includegraphics[width=\linewidth]{disk.jpg}}
   \end{minipage} \hspace{-0.75cm}&=\sum_{R,R'}\underbrace{\textrm{dim}(R)\chi_R(h_{AdS_{2}})}_{\text{for U(1)}}\underbrace{\textrm{dim}(R')\chi_{R'}(h_{AdS_2})}_{\text{for SO(3)}}\exp\Bigg\{4\pi(\varphi_0(Q))\Bigg\} \\ & \times \exp\Big(-\underbrace{\frac{\beta}{\epsilon}}_{\textrm{Divergent piece}}\varphi_0(Q)\Big)\sqrt{\frac{\varphi_{b,Q}^{3},\sqrt{U''(\varphi_0(Q))}\,}{\beta^3}}\,\, \exp({\frac{2\pi^2\varphi_{b,Q}}{\beta}})e^{-\beta \Tilde{M_0}(Q)}\,,\\&
   =\sum_{Q,j}\scriptstyle{\,\,\underbrace{e^{\beta \mu Q}}_{\text{for U(1)}}\,\,\underbrace{(2j+1)\,\chi_{j}(h_{AdS_2}^{SO(3)})}_{\text{for SO(3)$\equiv 1$ in $j=0$}}\exp\Bigg\{4\pi(\varphi_0(Q))\Bigg\} \sqrt{\frac{\varphi_{b,Q}^{3}\sqrt{U''(\varphi_0(Q))}\,}{\beta^3}}\,\exp({\frac{2\pi^2\varphi_{b,Q}}{\beta}})e^{-\beta \Tilde{M_0}(Q)}\,,} \\& 
   =\sum_{Q}\,\,\underbrace{e^{\beta \mu Q}}_{\text{for U(1)}}\,\,\underbrace{\exp\Bigg\{4\pi\varphi_0(Q)\Bigg\}}_{\equiv{S_0}(Q)\text{(charge dependent)}}\sqrt{\frac{\varphi_{b,Q}^{3}\sqrt{U''(\varphi_0(Q))}\,}{\beta^3}}\,\exp({\frac{2\pi^2\varphi_{b,Q}}{\beta}})\, e^{-\beta \Tilde{M_0}(Q)} \,, \\&
   =\sum_{Q}\,\,\underbrace{e^{\beta \mu Q}}_{\text{for U(1)}}\,e^{{S_0}(Q)}\,\sqrt{\frac{\varphi_{b,Q}^{3}\sqrt{U''(\varphi_0(Q))}\,}{\beta^3}}\,\exp({\frac{2\pi^2\varphi_{b,Q}}{\beta}})\,\underbrace{e^{-\beta \Tilde{M_0}(Q)}}_{\text{mass term}}\,. \label{4.12m}
\end{split}
\end{align}

\vspace{0.3cm}

We have assumed that we can independently integrate out the  the gauge degrees of freedom without affecting the measure for the Schwarzian field, $\mathcal{D}\mu[F]$. Also we have divided by the volume of the \textcolor{black}{$SL(2,R)$ group}  to avoid over-counting, as the isometry group of the hyperbolic disk is $SL(2,R)\,.$ Last but not the least,  $C_2(R')$ is not present as we have focused on the $j=0$ sector.

\vspace{0.6 cm}
Now, we proceed to compute the trumpet partition function. The trumpet is a genus zero surface with one asymptotic (nearly-) $AdS_2$ boundary and one geodesic boundary. The trumpet partition function has been evaluated below for the 2D gravity non-minimally coupled to the $U(1)$ and $SO(3)$ gauge field.\footnote{We do the computations of partition function for $j=0$.} \\\\
\textbf{ The trumpet partition function:}
We follow the same procedure as the $g=0$ disk partition function to calculate the trumpet partition function. The relevant geometry is obtained by a piece of hyperbolic space equipped with the following metric,
$$ds^2=d\xi^2+\cosh^2(\xi)d\tau^2\,\,\,\,\, \text{            with     } \,\,\,\,\,\,\,\tau \sim \tau + b\,.$$

For the trumpet, the hyperbolic plane's SL(2,R) symmetry breaks down to U(1) corresponding to the translation symmetry along $\tau$ direction. The wiggly boundary (as shown in the figure below) is parametrized by  $\tau(u)$ and the boundary action becomes, $-\varphi\int  \{e^{-\tau},u\}$. \textcolor{black}{Now, we can evaluate the boundary Gaussian integral as done in }\cite{Saad:2019lba}\,. Finally, we get using (\ref{3.13}) with Euler-characteristic $\chi_{\textrm{trumpet}}=0$ ,

\begin{align}
\centering
Z_{\textrm{\textcolor{black}{2DEMSO(3)}}}^{\textrm{trumpet}}\equiv\begin{minipage}[h]{0.12\linewidth}
	\vspace{4pt}
	\scalebox{1.1}{\includegraphics[width=\linewidth]{trumpet.jpg}}
   \end{minipage}\hspace{0.2 cm} = &\sum_{R,R'} \underbrace{\chi_R(h_{AdS_2})\chi_R(h_{b})\chi_{R'}(h_{AdS_2})\chi_{R'}(h_{b})}_{\text{Topological factor}}\\ & \times\underbrace{\sqrt{\frac{2\pi\,\sqrt{U''(\varphi_0(Q))}\varphi_{b,Q}}{\beta}}}_{\text{Contains Casimir}}\,\ \exp({\frac{-b^2\varphi_{b,Q}}{2\beta}})\exp\Big(-\underbrace{\frac{\beta}{\epsilon}}_{\textrm{Divergent piece}}\varphi_0(Q)\Big)\,,\nonumber\\&
   =\sum_{Q}\,\,\underbrace{e^{\beta \mu Q}}_{\text{for U(1)}}\,e^{-\beta \Tilde{M_0}(Q)}\,\sqrt{\frac{\sqrt{U''(\varphi_0(Q))}\varphi_{b,Q}}{\beta}}\exp({\frac{-b^2\varphi_{b,Q}}{2\beta}}) \times  \chi_{Q}(h_b)\chi_{j}(h_b)\,.\nonumber  
\end{align}

\vspace{0.5 cm}
Here the $h_{AdS_2}$ is the holonomy defined along the $AdS_2$ boundary, and $h_b$ is the holonomy along the geodesic boundary. $R, R'$ respectively denote representation of $U(1)$ and $SO(3)$ respectively.\\\\
\textbf{Relevant boundary condition:}
Before ending this section, we discuss the boundary condition on the $U(1)$ gauge field $A_{\mu}$ to remove any UV- divergences and make the path integral convergent. We are considering \textcolor{black}{2D} gravity coupled to U(1) and SO(3) gauge fields. The divergence term proportional to $\frac{1}{\epsilon^2}$ appearing from the extrinsic curvature can be regularized by adding a counter term proportional to $\frac{1}{\epsilon^2}\,.$ This has already been incorporated into the action itself. However, we need to change the boundary condition to regularise the divergence in action due to the presence of another divergent term $\sim\frac{\beta}{\epsilon}\,.$ Inside the path-integral, this term will appear in the exponential as $\exp\Big(\frac{-\beta}{\epsilon}(\varphi_0(Q))\Big)\,.$ This is removed by adding the defect action as mentioned in (\ref{eq 3.13}). This will further give rise to a term proportional to  $-Q^2$ in the exponential. This helps to converge the sum in the partition function.\\

Now the defect action, as stated in (\ref{eq 3.13}) leads to the modified boundary condition as follows,

$$A_u|_{\partial {\mathcal M}}-A_u|_I=-4i\sqrt g_{uu} \Bigg[\epsilon \Big({c} Q^2 \Big)-{\varphi_0(Q)}\Bigg]\Bigg|_{I}\,. $$

Hence, we get \begin{align}\delta\Bigg(A_{u}-4 i\sqrt g_{uu} \Bigg[\epsilon \Big({c} Q^2 \Big)-{\varphi_0(Q)}\Bigg] \Bigg)=0\,.\end{align}

\section{Brief review of $T\bar{T}$+$J\bar{T}$ deformation}\label{sec5}

One of the main goals of this paper is to compute the partition function of JT gravity coupled with $U(1)$ and $SO(3)$ deformed by  $T\bar{T}$ and $J\bar{T}\,.$ We briefly review some of the basic facts regarding these two deformations. \\

\textbf{$T\Bar{T}$ deformation:} The $T\bar{T}$ deformation is an integrable but irrelevant deformation. This composite operator was first studied by Zamolodchikov et al. and Tateo et al. \textcolor{black}{\cite{Smirnov:2016lqw,Cavaglia:2016oda}}. It is constructed from an energy-momentum tensor in two-dimensional QFT. The $T\bar{T}$ deformation is a special case of more generic irrelevant integrable deformations of integrable QFTs (IQFT) introduced in \cite{Smirnov:2016lqw}. 
The composite operator $T\bar{T}$ in two-dimensional quantum field theory is constructed from the chiral components $T$ and $\bar{T}$ of the energy-momentum tensor $T_{\mu\nu}$.
Labelling the complex coordinates as z,$\bar z$ we can write the chiral components of the energy-momentum tensor as follows,
\begin{equation}
   T= -2\pi T_{zz},\,\,\,\,\,\,\,\,\, \bar T= -2\pi T_{\bar z\bar z},\,\,\,\,\,\,\,\,\, \Theta =2 \pi T_{z\bar z}\,.
\end{equation}

This yields the following,
\begin{equation}\langle T \bar T\rangle= \langle T\rangle \langle  \bar T\rangle-\langle\Theta\rangle^2\,.\end{equation}
In the limit $z\rightarrow z'$, \,  $T(z)\bar{T}(z')$ and $\Theta(z)\bar{\Theta}(z')$ both are divergent. But for the combination $T(z)\bar{T}(z')$ - $\Theta(z){\Theta}(z')$  the divergent parts cancel each other. Therefore, this specific combination is well defined up to total derivative terms.  Putting the theory on the cylinder, the expectation value can be computed as :
\begin{equation}
    \langle n |T\bar T|n\rangle =  \langle n |T(z)|n\rangle \langle n|\bar T(z')|n\rangle - \langle n |\Theta(z)|n\rangle \langle n| \Theta(z')|n \rangle  \label{eq 4.3}
\end{equation}   
Using the fact,
$$T_{xx}=\frac{-1}{2\pi}(\bar T+T-2\Theta),\,\,\,\,\,\,\,\,
T_{yy}=\frac{1}{2\pi}(\bar T+T+2\Theta),\,\,\,\,\,\,\,\,
T_{xy}=\frac{i}{2\pi}(\bar T-T)$$ we can write (\ref{eq 4.3}) as,

\begin{align}
    \langle n |T\bar T|n\rangle =-\pi^2 ( \langle n |T_{xx}|n\rangle \langle n|\bar T_{yy}|n\rangle - \langle n |T_{xy}|n\rangle \langle n| T_{xy}|n \rangle)\,.  \label{5.4}
\end{align}   
Here, $T_{xy}$ is the momentum density, and $T_{yy}$ is the energy density considering $y$ as the direction of time.
Now, $ \langle n| T_{yy} |n\rangle = -\frac{1}{R}E_n(R)$ and $ \langle n| T_{xy} |n\rangle = -\frac{i}{R}P_n(R)\,,$ where the momentum $P_n(R)=\frac{2\pi l_n}{R}$, with $l_n$ being an integer. The partition function $Z$ is given by,
$$Z=\sum_n e^{-\beta E_n(R)+\mu q(n)}\,,$$
where $\mu =\mu(n)$ is the chemical potential. Now 
\begin{align}\langle n |T_{xx}| n\rangle=\frac{\delta}{\delta g^{xx}} Z\,.\end{align} 
Then, varying $R$,(radius of the cylinder) we get,
$$\frac{d}{dR}Z=-\beta \sum_n \frac{dE_n}{dR} e^{-\beta E_n(R)+\mu q(n)}\,.$$

With these in hand, the expression in (\ref{5.4}) can be recasted as,
\begin{align}
   \langle n |T\bar{T}| n\rangle = -\frac{\pi^2}{R}\Bigg( E_n(R)\frac{d}{dR} E_n(R)+\frac{P_n^2(R)}{R}\Bigg)\,.
\end{align}\\
\textbf{$J\Bar{T}$ deformation:} Another important irrelevant deformation of the Smirnov-Zamolodchikov class to 2D CFT is the $J\Bar{T}$-deformation, which has the following form:
\begin{align}
    \begin{split}
S_{\text{def.}}=S_{\text{CFT}}+\alpha\int d^2 z \, J \Bar{T}
    \end{split}
\end{align}
where,  $\bar{T}$ is the anti-holomorphic part of stress-tensor and $J$ is a $U(1)$ current. $J\bar{T}$  breaks Lorentz symmetry of the theory, unlike $T\Bar{T}$ deformation. Using the above facts, one can construct the deformed spectrum of the theory as \cite{Chakraborty:2018vja, Guica:2017lia, Chakraborty:2020xwo}.
\begin{align}
    \begin{split}
        E(\alpha,E_{0},P,Q)=P+\frac{2}{\alpha^2}\Bigg[\Big(1-\alpha Q\Big)-\sqrt{\Big(1-\alpha Q\Big)^2-\alpha^2 (E_0-P)}\Bigg].\label{5.8}
    \end{split}
\end{align}
where, $E_0$ and $Q$ are the undeformed energy and $U(1)$ charge and $P$ is the momentum. Also, we assume that along the flow there is no deformation of the charge parameter $Q$. For, quantum mechanical systems, (i.e. in $0+1D$) $P=0$  %and subsequently, the deformed spectrum in (\ref{5.8}) reduces to (\ref{6.2}) with $\lambda=0$. 
In the next section, we will consider the linear combination of $T\Bar{T}$ and $J\Bar{T}$ deformation in quantum mechanics and discuss the computation of the partition function, which will be relevant for our subsequent computations.

\section{$T\bar{T}$+$J\bar{T}$ deformed correlation function in $U(1)$ coupled JT-gravity} \label{sec6}
In Sec.~(\ref{section 3}), we have computed the partition function of the undeformed theory. In this section, we will consider the $T\bar{T}$+$J\bar{T}$ deformation and compute the correlation function. First, we will discuss the $T\bar{T}$+$J\bar{T}$ deformation in the context of quantum mechanics. Next, we will discuss the effect of this deformation, particularly on the \textcolor{black}{one-point and two-point} functions of partition function.\\

\textbf{$T\bar{T}$+$J\bar{T}$ deformation in quantum mechanics:} $T\bar{T}$ and $J\bar{T}$ deformation in a generic quantum mechanical system coupled with U(1) charge can be written as the following flow equations \cite{Gross:2019ach},
\begin{align}
    \begin{split}
      &  \frac{\partial \mathcal{L}_{\lambda,\alpha}}{\partial \lambda}=-\frac{1}{2}T\bar{T}(\lambda)\,,\\ &
      \frac{\partial \mathcal{L}_{\lambda,\alpha}}{\partial \alpha}=2J\bar{T}(\alpha)\,.
    \end{split}
.\end{align}
Here, $\alpha$ and $\lambda$ are the deformation parameters for the $J\bar{T}$ and $T\bar{T}$ deformations, respectively. The deformed spectrum is given by,
\begin{align}
    \begin{split}
        E(\alpha,\lambda)=-\frac{2}{2\lambda-\alpha^2}\Big[1-\alpha Q-\sqrt{(1-\alpha Q)^2+E_0 (2\lambda-\alpha^2)}\Big]\,.\label{6.2}
    \end{split}
\end{align}

We define \textcolor{black}{\cite{Chakraborty:2020xwo}} $$h\equiv\frac{1}{2\lambda-\alpha^2}\,.$$
Here, $Q$ and $E_0$ are the undeformed charge and energy, and also we assume that the charge $Q$ remains undeformed along the flow. The kernel for this transformation, which will be useful later for computing the deformed partition function, is given by
\begin{align}
    \begin{split} \label{kernel}
K(\alpha,\lambda,\beta,\beta',\mu,\mu')=\frac{\beta}{ \beta'^2\alpha}\exp\,\Big(\frac{(\mu'-\mu)^2}{4h\alpha^2\beta'}-\frac{(\beta'-\beta)}{\alpha\beta'}{(\mu'-\mu)}\Big)\,.
\end{split}
\end{align}
 \textit{$T\bar{T}$ and the $J\bar{T}$ deformation breaks certain symmetries (e.g. scale invariance) of the underlying theory. It is likely that when $h\rightarrow \infty$, some of the broken symmetries are restored. This is reflected in the deformed 2-point function, which `looks almost undeformed" in this limit with $\mu =0$}. However, in the presence of some non-vanishing chemical potential, which is the case for us, there is a certain change in the two-point function.

Using the kernel mentioned in (\ref{kernel}), we can relate the undeformed partition function with the deformed one using the following relation \cite{Ebert:2022gyn}.

\begin{align}
\begin{split}
    \langle{Z}(\beta_1,\mu_1)\cdots {Z}(\beta_n,\mu_n)\rangle_{\alpha,\lambda}=&%\int d\beta_1'd\mu'_1K(\alpha,\lambda,\beta,\beta',\mu,\mu')\cdots \int d\beta_n' \,d\mu_{n}'\,K(\alpha,\lambda,\beta,\beta',\mu,\mu')\\&\hspace{5.6 cm}\langle\mathcal{Z}(\beta_1',\mu_1')\cdots \mathcal{Z}(\beta_n',\mu_n')\rangle_0\\ &
    =\prod_{i=1}^{n}\int_{0}^{\infty}d\beta_{i}'\int_{-i \infty}^{i \infty}  d\mu_{i}' \,K(\alpha, \lambda, \beta_{i}, \beta_{i}', \mu_{i}, \mu_{i}')\,\langle {Z}(\beta_1',\mu_1')\cdots {Z}(\beta_n',\mu_n')\rangle_0\,.
\end{split}
\end{align}
$h$ is usually greater than zero for the sake of the convergence of the integral. However, it can be analytically continued to $h<0$ case.
% Usually in JT gravity deforming the Schwarzian theory with $T\bar{T}$ deformation is same as adding a bulk cut-off with deformed boundary conditions, which can also be explained as a new way of stating the variational principle well-defined, where a linear combination of undeformed source and dual expectation value is held fixed, and this combination can be interpreted as the deformed source
\noindent

\textcolor{black}{Now, we proceed to calculate the deformed 2D gravity partition function by applying the $T\bar{T}$+$J\bar{T}$ deformation of the (0+1) dimensional boundary theory. The computation of the bulk partition function in JT gravity is done by integrating the $S[f]$ (as defined in (\ref{schwarz}) in the path integral over the bulk moduli as demonstrated in the section~(\ref{sec4}) for disk and trumpet. Furthermore, it was shown in \cite{Ebert:2022ehb, Chakraborty:2020xwo}, that the flow equation of the deformed ($T\bar{T}+J\bar{T}$) theory in the boundary matches with the flow equation of the thermal partition function in the bulk. So, we use the same deformation kernel (used for deforming the boundary Schwarzian theory) to deform the bulk partition function in $AdS_2$. For a more detailed explanation, we refer the reader to \cite{Ebert:2022ehb, Chakraborty:2020xwo}. One should note that we are not deforming the bulk theory of JT gravity by applying the $T\bar{T}+J\bar{T}$ deformation. Rather, we deform the boundary dual by a family of integrable deformation, which changes the ultraviolet behaviour of the theory. Now we have the deformed partition function as follows,}
\begin{align}
\begin{split}
    \langle{Z}(\beta_1,\mu_1)\rangle_{\alpha,\lambda}=&\int d\beta_1'd\mu'_1K(\alpha,\lambda,\beta,\beta',\mu,\mu')\langle{Z}(\beta_1',\mu_1')\rangle_0\,.
    \label{3.5}
\end{split}
\end{align}
\textcolor{black}{At this point we would like to make some comment on the above prescription for the computation of deformed JT gravity partition function. Although we are using the term `deformed JT', this terminology is quite misleading. As of, we are not deforming the bulk gravity theory rather we are deforming the boundary quantum mechanical theory and want to investigate the effect of deforming the boundary theory to the bulk observables.}
As before, the one-point partition function of the undeformed theory is given by,
$$
\langle\mathcal{Z}(\beta_1',\mu_1')\rangle_0 =Z_{\textcolor{black}{\textrm{2DEMSO(3)}}}^{\textrm{disk}}+\sum_g\int d\Tilde h\int b\, db \,Z_{\textrm{2DEMSO(3)}}^{\textrm{trumpet}}Z_{\textcolor{black}{\textrm{2DEMSO(3)}}}^{(g,1)}\,.$$

Now, using (\ref{3.5}), we calculate the deformed one-point partition function \footnote{we use the fact $\int d\Tilde{h}\, \chi_R^2(h_b)=1$ in the following sections during calculation. }, which has two parts, coming from the disk and the trumpet geometry.
The part coming from the disk geometry after the kernel integration is basically given by: \begin{align}
\begin{split}\mathcal{I}^{\textrm{disk}}&=\int_0^{\infty} d \beta'\int_{-i \infty }^{i \infty } d\mu'\frac{\beta}{\beta'^2\alpha}\exp\,(\frac{(\mu'-\mu)^2}{4h\alpha^2\beta'}-\frac{(\beta'-\beta)}{\alpha\beta'}{(\mu'-\mu)}) \frac{\varphi_{b,Q}^{3/2}\,U''(\varphi_0(Q))^{1/4}}{\sqrt{2\pi }{\beta'}^{3/2}}\,\\& \times e^{\frac{2\pi^2\varphi_{b,Q}}{\beta'}+\beta' \mu' Q-\beta' \Tilde{M_0}(Q)}\,.\label{6.6}\end{split}\end{align}
Though this integral looks a bit messy, it is not hard to see that in the regime  $h\rightarrow\infty$  (unitary \cite{Chakraborty:2020xwo})  one can first perform the inverse Laplace transform of the remaining part except for the non-linear($\sim \mu^2$) term in the exponential, which gives the following delta function,
\begin{align}
    \int_{-i \infty }^{i \infty } d\mu' \exp\,(-\frac{(\beta'-\beta)}{\alpha\beta'}{(\mu'-\mu)}+\beta' \mu' Q)e^{-\beta' \Tilde{M_0}(Q)}=\delta(\frac{\beta-\beta'}{\alpha \beta'}+\beta'Q)\exp{\frac{(\beta'-\beta)\mu}{\alpha \beta'}}e^{-\beta' \Tilde{M_0}(Q)}\,.
    \label{3.8}
\end{align}
%This above line comes from the \textit{Inverse Laplace transformation.}
The delta function can be massaged in the following way,
\begin{align}\delta(f(x))=\frac{1}{|f'(x)|_{x_0}}\sum \delta(x-x_0)\,.\end{align}
For our case $$f'(x)\equiv 1-2\alpha Q x\,.$$
Hence, (\ref{3.8}) up to $\mathcal{O}(\frac{1}{h^0})$ can be written as :
\begin{align}
    \begin{split}   
   \delta(\frac{\beta-\beta'}{\alpha \beta'}+\beta'Q)exp{\frac{(\beta'-\beta)\mu}{\alpha \beta'}}e^{-\beta' \Tilde{M_0}(Q)}=& \alpha\beta'\delta(\alpha\beta'^2 Q-\beta'+\beta)exp{\frac{(\beta'-\beta)\mu}{\alpha \beta'}}e^{-\beta' \Tilde{M_0}(Q)}\,,\\&=\alpha\beta'\Big\{\frac{\delta(\beta'-\Delta)}{|f'(\Delta)|}+\frac{\delta(\beta'-\xi)}{|f'(\xi)|}\Big\}exp{\frac{(\beta'-\beta)\mu}{\alpha \beta'}} e^{-\beta' \Tilde{M_0}(Q)}\,,\end{split}
\end{align}

where $\Delta$ and $\xi$ are two zeros of the quadratic function.
Now we do the $\beta'$ integral along the real axis. This gives the following:
 \begin{align}
    \begin{split}
\mathcal{I}^{\textrm{disk}}=&\int_0^{\infty}d\beta'\,\frac{\beta{\varphi_{b,Q}^{3/2}\,U''(\varphi_0(Q))^{1/4}}}{{(2\pi)}^{1/2} \beta'^{5/2}}\exp{\frac{(\beta'-\beta)\mu}{\alpha \beta'}} \exp{\frac{2\pi^2\varphi_{b,Q}}{\beta'}}\Big\{\frac{\delta(\beta'-\Delta)}{|f'(\Delta)|}+\frac{\delta(\beta'-\xi)}{|f'(\xi)|}\Big\}e^{-\beta' \Tilde{M_0}(Q)}\,,\\&
=\Bigg[\frac{\beta{\varphi_{b,Q}^{3/2}\,\,U''(\varphi_0(Q))^{1/4}}}{{(2\pi)}^{1/2} \Delta^{5/2}}exp{\frac{(\Delta-\beta)\mu}{\alpha \Delta}}\exp{\frac{2\pi^2\varphi_{b,Q}}{\Delta}}\frac{1}{|f'(\Delta)|}\Bigg]e^{-\Delta \Tilde{M_0}(Q)}+(\Delta\longleftrightarrow \xi)\,.
    \end{split}
\end{align}
The two zeros of the quadratic function $f(\beta')$ are defined as follows:
\begin{align}
\begin{split}
\Delta:=\frac{1+\sqrt{1-4\alpha Q\beta }}{2\alpha Q} \text{  and  } \xi:=\frac{1-\sqrt{1-4\alpha Q\beta }}{2\alpha Q}\,.\label{7.4e}
\end{split}
\end{align}

The expressions of $\Delta$ and $\xi$ in (\ref{7.4e}) put a constraint on the upper limit ($Q_{max}$) tosummation over charge in the path-integral. It is clear from the expression that to keep $\Delta$ and $\xi$ real, we need to make the term under the square root positive.\par

Next, to compute the higher genus one-point function, we focus on the $\mathcal{V}_{g,1}$ for $g\gg 1$ limit. 
We obtain the one point function (upto genus one) as,

\begin{align}
\begin{split}
    \Big\langle{Z}&(\beta,\mu)\Big\rangle_{\alpha,\lambda,g=1}=\sum_Q\Bigg\{\Bigg[\frac{\beta{\varphi_{b,Q}^{3/2}\,U''(\varphi_0(Q))^{1/4}}}{{(2\pi)}^{1/2} \Delta^{5/2}}\exp{\frac{(\Delta-\beta)\mu}{\alpha \Delta}}\exp{\frac{2\pi^2\varphi_{b,Q}}{\Delta}}\frac{1}{|f'(\Delta)|} \Bigg]e^{-\Delta \Tilde{M_0}(Q)}e^{S_0(Q)}\\&+\scriptstyle{(\Delta\longleftrightarrow \xi)+\Bigg[\frac{\beta e^{-\Delta \Tilde{M_0}(Q)}}{ \Delta^{3/2} |f'(\Delta)|}\exp\frac{(\Delta-\beta)\mu}{\alpha \Delta }\frac{\Delta  \left(\Delta  +2 \pi ^2 \varphi \right)}{24 \varphi ^2}+\frac{\beta e^{-\xi \Tilde{M_0}(Q)}}{\xi^{3/2} |f'(\xi)|}\exp\frac{(\xi-\beta)\mu}{\alpha \xi }\frac{\xi  \left(\xi +2 \pi ^2 \varphi_{b,Q} \right)}{24 \varphi_{b,Q} ^2}\Bigg]\sqrt{{\varphi_{b,Q}U''(\varphi_0(Q))}}}e^{-S_0(Q)}\Bigg\}\,.
   \label{4.13}
   \end{split}
\end{align}
where, $\mathcal{V}_{g,1}(b)$ is the Weil-Petersson volume of the moduli space of hyperbolic Riemann surfaces with genus `$g$' and one geodesic boundary of length `$b$'. The details of integration are given in the Appendix (\ref{a2})\,.\\\\
\textbf{Higher genus contribution:} Assuming $g\gg b$ the volume of the moduli space is given by \cite{Saad:2019lba},
\begin{align}
    \begin{split}
        \mathcal{V}_{g,1}(b)\approx \frac{4(4\pi^2)^{2g-\frac{3}{2}}}{(2\pi)^{3/2}}\Gamma(2g-\frac{3}{2})\frac{\sinh{b/2}}{b}\,.
    \end{split}\label{2.19}
\end{align}

Some of the details of the calculation one-point function considering a higher genus contribution have been given in Appendix (\ref{B1}) and Appendix (\ref{B.2}).\par

Now, to calculate the two-point connected partition function, we glue the no-handled trumpet with another no-handle trumpet via the moduli space containing the $g$-genus.
This will give the expression of $Z_{g,2}$. Using (\ref{2.9}), for the deformed case,
\begin{align}
    \langle Z(\beta_1,\mu_1) Z(\beta_2,\mu_2)\rangle_{\text{conn}}=Z_{0,2}(\beta_1,\mu_1;\beta_2,\mu_2)+\sum_{g=1}^{\infty}e^{-2g S_0}Z_{g,2}(\beta_1,\mu_1;\beta_2,\mu_2)\,.
    \label{6.11}
\end{align}
Hence $Z_{g,2} $ in Fig~(\ref{fig 1}) can be written as,

\vspace{-0.45 cm}
\begin{align}Z_{g,2}&= {\sum_{R,R'} \text{dim(R)}^{\chi_L}\chi_R(h_{AdS_2})\chi_R(h_{b})\text{dim}(R')^{\chi_L}}\\&{\chi_{R'}(h_{AdS_2})\chi_{R'}(h_{b})\int_{0}^{\infty}  b_1\, db_1\, b_2\, db_2\,\,\mathcal{V}_{g,2}(b_1,b_2)Z_{2DEMSO(3)}^{trumpet}(b_2)\,\,\sqrt{\frac{\varphi_{b,Q}}{2\pi\beta }}\,\,U''(\varphi_0(Q))^{1/4} e^{\frac{-b_1^2\varphi_{b,Q}}{2\beta}}}e^{-\beta \Tilde{M_0}(Q)}\,.\nonumber\end{align}

\begin{figure}
\begin{center}
\scalebox{0.15}{\includegraphics{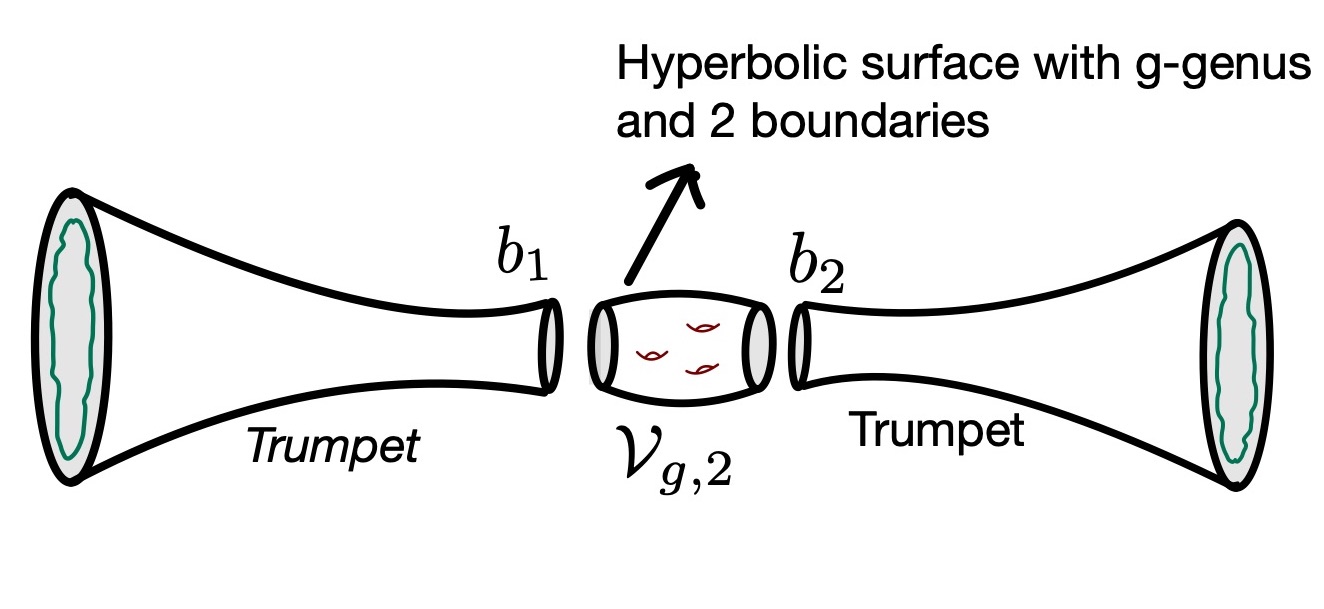}}
\end{center}
\caption{Gluing two trumpets with different genus structure to find $Z_{g,2}$}
\label{fig 1}
\end{figure}
Here $\chi_L=-2g$.
Therefore, the two-point function can be written directly as:
\begin{align}Z_{0,2}&=\int_0^\infty db\, b \, Z_{2DEMSO(3)}^{trumpet}(\beta_1,\mu_1)Z_{2DEMSO(3)}^{trumpet}(\beta_2,\mu_2)\end{align}

In (\ref{6.11}),  the first term is the \textit{genus 0} wormhole contribution, and the other terms consist of higher genus wormholes. The double trumpet geometry is mainly responsible for the \textit{ramp} in the spectral form factor, which we will discuss in Section \ref{sec8}.\footnote{We also discussed the ramp structure from the perspective of late time topology change in \textcolor{black}{2DEM} theory.} Putting all these together, we finally get
\vspace{-0.3 cm}
\begin{align}
\begin{split}
 \Big\langle Z(\beta_1,\mu_1) Z(\beta_2,\mu_2)\Big\rangle_{\text{conn}}=&
 \underbrace{\int_0^\infty b db Z_{\textcolor{black}{\textrm{2DEMSO(3)}}}^{\textrm{trumpet}}(\beta_1,\mu_1)Z_{\textcolor{black}{\textrm{2DEMSO(3)}}}^{\textrm{trumpet}}(\beta_2,\mu_2)}_{\mathcal{I}_1}\Bigg(\equiv \begin{minipage}[h]{0.12\linewidth}
	\vspace{4pt}
	\scalebox{1.2}{\includegraphics[width=\linewidth]{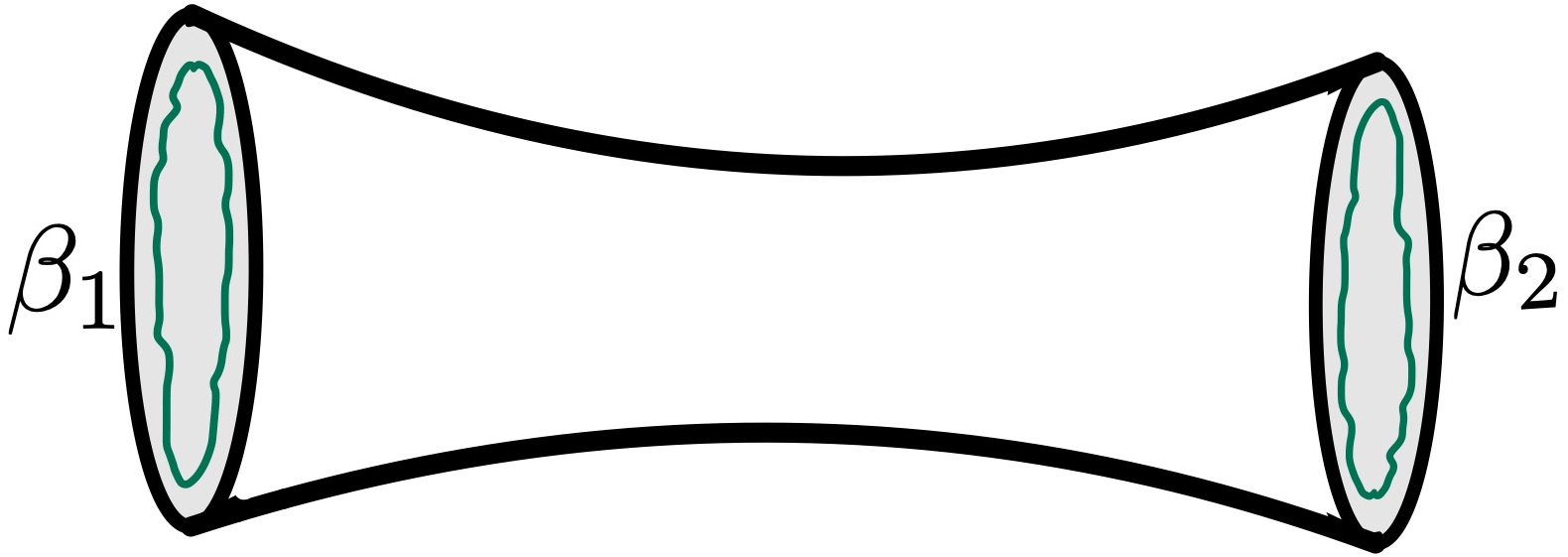}}
   \end{minipage}\ \hspace{0.3 cm}\Bigg)\\&+\sum_{g=1}^{\infty}e^{-2gS_0(Q)}\int_0^\infty  b_1 db_1  b_2 db_2\sum_{R,R'} \text{dim}(R)^{\chi_L}\chi_R(h_{AdS_2})\chi_R(h_{b})\text{dim}(R')^{\chi_L}\\&\underbrace{\chi_{R'}(h_{AdS_2})\chi_{R'}(h_{b})   \,\mathcal{V}_{g,2}(b)Z_{2DEMSO(3)}^{trumpet}\,\,\sqrt{\frac{\varphi_{b,Q}}{2\pi\beta }}\,U''(\varphi_0(Q))^{1/4} e^{\frac{-b^2\varphi_{b,Q}}{2\beta}}e^{-\beta \Tilde{M_0}(Q)}}_{\mathcal{I}_2}\,.
    \label{3.14}
    \end{split}
\end{align} 
%\vspace{-0.8 cm}
The first integral $\mathcal{I}_1$ in (\ref{3.14}) can be computed as follows.
\begin{align}
\begin{split}
\mathcal{I}_1&=\int_0^\infty b db \sqrt{\frac{\varphi_{b,Q}}{2\pi\beta_1}}\, e^{\frac{-b^2\varphi_{b,Q}}{2\beta_1}} \sqrt{\frac{\varphi_{b,Q}}{2\pi\beta_2}}\,U''(\varphi_0(Q))^{1/2} e^{\frac{-b^2\varphi_{b,Q}}{2\beta_2}} e^{-\beta_1 \Tilde{M_0}(Q)} e^{-\beta_2 \Tilde{M_0}(Q)}\\ &\hspace{2 cm}\times\underbrace{\textrm{(other factors)}}_{\text{$\mathcal{N} \equiv$ Topological term $\times$ Casimir}}\,,\\& 
=\frac{\mathcal{N} \,U''(\varphi_0(Q))^{1/2}\varphi_{b,Q}}{2\pi \sqrt{\beta_1\beta_2}}\int_0^\infty db \,b \,e^{\frac{-b^2\varphi_{b,Q}}{2\beta_1}-\frac{b^2\varphi_{b,Q}}{2\beta_2}} e^{-\beta_1 \Tilde{M_0}(Q)} e^{-\beta_2 \Tilde{M_0}(Q)}\,,\\&
=\frac{\mathcal{N}  \,U''(\varphi_0(Q))^{1/2}\sqrt{\beta_1\beta_2}}{2\pi(\beta_1+\beta_2)}e^{-\beta_1 \Tilde{M_0}(Q)} e^{-\beta_2 \Tilde{M_0}(Q)}\label{6.18m}\,.
\end{split}
\end{align}
Within these other factors in (\ref{6.18m}) mentioned above, we have some $\beta$ and $\mu$ dependence coming from the group character and other terms. Similarly, the second one $\mathcal{I}_2$ in (\ref{3.14}) can be written as,
{\begin{align}
\begin{split}
    \mathcal{I}_2&=\sum_{g= 1}^\infty \frac{\mathcal{N}}{U''(\varphi_0(Q))^{-3/4}}  \exp(-2gS_0(Q))\int_0^\infty db_1\, db_2\, b_1\, b_2\, \mathcal{V}_{g,2}(b_1,b_2)Z_{2DEMSO(3)}^{Trumpet}\scriptstyle{\sqrt{\frac{\varphi_{b,Q}}{2\pi\beta_1}}\, e^{\frac{-b_1^2\varphi_{b,Q}}{2\beta_1}}e^{-\beta_1 \Tilde{M_0}(Q)} e^{-\beta_2 \Tilde{M_0}(Q)}\,,}\\&
  %  \approx \sum_{g\gg 1}^\infty exp(-2gS_0)\int_0^\infty db b Z_{2DEMSO(3)}^{Trumpet}\sqrt{\frac{\varphi_{b,Q}}{2\pi\beta}_1}\, e^{\frac{-b^2\varphi_{b,Q}}{2\beta_1}}(other factors)\\&
 %   \approx \sum_{g\gg 1}^\infty exp(-2gS_0)\frac{4(4\pi^2)^{2g-\frac{3}{2}}}{(2\pi)^{3/2}}\Gamma(2g-\frac{3}{2})\frac{\varphi_{b,Q}}{2\pi(\sqrt{\beta_1\beta_2})}\int _0^\infty db \, b\, e^{\frac{-b^2\varphi_{b,Q}}{2\beta_1}}e^{\frac{-b^2\varphi_{b,Q}}{2\beta_2}}\frac{\sinh{b/2}}{b}(other factors)\\&
  %  \approx  \sum_{g\gg 1}^\infty exp(-2gS_0)\frac{4(4\pi^2)^{2g-\frac{3}{2}}}{(2\pi)^{3/2}}\Gamma(2g-\frac{3}{2})\frac{\varphi_{b,Q}}{2\pi(\sqrt{\beta_1\beta_2})}\frac{\sqrt{\frac{\pi }{2}} e^{\frac{\text{$\beta $1} \text{$\beta $2}}{8 \text{$\beta $1} \varphi_{b,Q} +8 \text{$\beta $2} \varphi_{b,Q} }} \text{erf}\left(\frac{1}{2 \sqrt{2} \sqrt{\varphi_{b,Q}  \left(\frac{1}{\text{$\beta $1}}+\frac{1}{\text{$\beta $2}}\right)}}\right)}{\sqrt{\varphi_{b,Q}  \left(\frac{1}{\text{$\beta $1}}+\frac{1}{\text{$\beta $2}}\right)}}(other factors)\\&
   % \approx \sum_{g\gg 1}^\infty exp(-2gS_0)\frac{4(4\pi^2)^{2g-\frac{3}{2}}}{(2\pi)^{3/2}}\Gamma(2g-\frac{3}{2})\frac{\sqrt{\varphi_{b,Q}}}{2\pi}\frac{\sqrt{\frac{\pi }{2}} e^{\frac{\text{$\beta $1} \text{$\beta $2}}{8 \text{$\beta $1} \varphi_{b,Q} +8 \text{$\beta $2} \varphi_{b,Q} }} \text{erf}\left(\frac{1}{2 \sqrt{2} \sqrt{\varphi_{b,Q}  \left(\frac{1}{\text{$\beta $1}}+\frac{1}{\text{$\beta $2}}\right)}}\right)}{\sqrt{ \left({\text{$\beta $1}}+{\text{$\beta $2}}\right)}}(other factors)
\end{split}
\end{align} 
}
In the above sum we only focus on the genus one contribution:
\begin{align}
\begin{split}
\mathcal{I}_{2}&\xrightarrow[g=1] {}
\frac{\mathcal{N}}{U''(\varphi_0(Q))^{-3/4}} \frac{\exp(-2 S_0(Q))}{\textcolor{black}{192}} \int_0^\infty e^{\frac{-b_1^2\varphi_{b,Q}}{2\beta_1}} e^{\frac{-b_2^2\varphi_{b,Q}}{2\beta_2}} db_1 db_2 b_1 b_2{(4\pi^2+b_1^2+b_2^2)(12\pi^2+b_1^2+b_2^2)} \\ & \hspace{5 cm}\times
\sqrt{\frac{\varphi_{b,Q}}{2\pi\beta_1}}\sqrt{\frac{\varphi_{b,Q}}{2\pi\beta_2}}e^{-\beta_1 \Tilde{M_0}(Q)} e^{-\beta_2 \Tilde{M_0}(Q)}\,,\\&
    =\frac{\mathcal{N}}{U''(\varphi_0(Q))^{-3/4}}  \exp(-2 S_0(Q)) \sqrt{\frac{\varphi_{b,Q}}{2\pi\beta_1}}\sqrt{\frac{\varphi_{b,Q}}{2\pi\beta_2}} \frac{\beta _1 \beta _2 \left[4 \pi ^2 \left(\beta _1+\beta _2\right) \varphi _b+6 \pi ^4 \varphi _b^2+(\beta_1+\beta_2)^2\right]}{24 \varphi_{b,Q}^4}\\ &\hspace{5 cm}\times e^{-\beta_1 \Tilde{M_0}(Q)} e^{-\beta_2 \Tilde{M_0}(Q)}\,.
    \end{split}
\end{align}
Then finally adding these two we get the connected undeformed two-point partition function, 
\begin{align}
\begin{split}
   \Big\langle Z(\beta_1,\mu_1) Z(\beta_2,\mu_2)\Big\rangle\Big |_{g=1,\textrm{undef.
}}&=\frac{\mathcal{N}}{U''(\varphi_0(Q))^{-3/4}} \exp(-2 S_0(Q)) \frac{\varphi_{b,Q}}{2\pi}\scriptstyle{\frac{\sqrt{\beta _1 \beta _2} \left[4 \pi ^2 \left(\beta _1+\beta _2\right) \varphi _b+6 \pi ^4 \varphi _b^2+(\beta_1+\beta_2)^2\right]}{24 \varphi _b^4}}\\ &\times  e^{-\beta_1 \Tilde{M_0}(Q)} e^{-\beta_2 \Tilde{M_0}(Q)}+\frac{\mathcal{N} \sqrt{ U''(\varphi_0(Q)) }\sqrt{\beta_1\beta_2}}{2\pi(\beta_1+\beta_2)}e^{-\beta_1 \Tilde{M_0}(Q)} e^{-\beta_2 \Tilde{M_0}(Q)}\,.
\end{split}
\end{align}
%\vspace{0.5 cm}
It contains the contribution from the \textit{g=0 trumpet} and  \textit{g=1 trumpet}. This undeformed two-point function can be used to calculate the deformed two-point function using (\ref{3.5})\footnote{We do  not show it explicitly as expression is long and not so illuminating.}.
To ensure that the metric remains hyperbolic (Euler Character becomes negative), the number of genus must satisfy $g >> 2$. But in our case, the Euler characteristic is already negative because of the presence of a one-boundary component of the moduli space.

\section{Annealed and Quenched Free Energy}\label{sec7}
For a random Hamiltonian \{$H$\} system, one can compute in general two different kinds of free energy as discussed in Sec.~(\ref{intro}): ``Quenched" ($F_{q}=-T\langle\log Z\rangle$) and ``Annealed" ($F_{a}=-T\log\langle Z \rangle$) where $Z(\beta)=\text{Tr}\,e^{-\beta H}$ is the partition function of theory and $\beta$ is the inverse temperature. The angular bracket $\langle \cdot\rangle$ denotes the ensemble average. In a random system, the thermodynamic properties are encoded into the quenched free energy: $F_{q}=-T\langle\log Z\rangle$ and entropy: $S=-\frac{\partial F_{q}}{\partial T}$.
The entropy is supposed to satisfy the following inequality: $S\ge 0$, implying that the quenched free energy should be a monotonically decreasing function of $T$. However, the annealed free energy $F_{a}$ is not necessarily a monotonic function of $T$. The computation of quenched free energy involves the computation of $\langle \log Z \rangle$, which can be calculated by the replica trick as,

\begin{align}
    \begin{split}
        \langle \log Z \rangle= \lim_{n\rightarrow 0}\frac{\langle Z^n \rangle-1}{n}\,. \label{7.1}
    \end{split}
\end{align}
However, the computation, especially taking the limit $n\rightarrow 0,$ of quenched free energy using the replica trick (\ref{7.1}) is complicated. For our case, it is useful to use the integral representation of the quenched free energy following \cite{Okuyama:2021pkf}.

\begin{align}
    \begin{split}
     {F}_{q}=-T \langle \log {Z}(\beta,\mu)\rangle_{\alpha,\lambda}=-T\Bigg[\log\langle {Z}(\beta,\mu)\rangle_{\alpha,\lambda}-\int_0^\infty \frac{dx}{x}\Bigg(\langle e^{-Zx} \rangle-e^{-\langle Z \rangle x}\Bigg)\Bigg].\label{7.2m}
    \end{split}
\end{align}
Now introducing the generating function $ \mathcal{Z}(x) $ of the connected correlator $\langle Z^n\rangle_c$
where,
\begin{equation} 
\mathcal{Z}(x)=\sum_{n=1}^\infty \frac{(-x)^n}{n!}\langle Z^n\rangle_c
\end{equation}

Using $\langle e^{-zx}\rangle=e^{-\mathcal{Z}(x)}$ we can write the equation (\ref{7.2m}) following \cite{Okuyama:2021pkf}\,,
%\vspace{-0.5cm}
\begin{align}
     -\frac{{F}_{q}(T)}{T}=\langle \log {Z}(\beta,\mu)\rangle_{\alpha,\lambda}=\log\langle {Z}(\beta,\mu)\rangle_{\alpha,\lambda}-\int_0^\infty \frac{dx}{x}\Bigg[e^{-\mathcal{Z}(x)}-e^{-\langle Z(\beta,\mu) \rangle x}\Bigg]\,.
     \label{4.17}
\end{align}
\par
For small temperature ($T\le e^{-2S_0(Q)/3}$) and small chemical potential such that $\mu\beta\ll 1$ \cite{Okuyama:2021pkf,Ebert:2022gyn}, \footnote{One can motivate this, by observing that,$\frac{\langle Z(\beta_1,\mu_1) Z(\beta_2,\mu_2)\rangle}{\langle Z(2\beta,2\mu)\rangle}\Big |_{\beta_{1,2}=\beta, \mu_{1,2}=\mu}\approx 1$ in this limit. However, the check for higher point correlations is numerically hard.}  $$\langle  Z(\beta,\mu)^n\rangle_c\approx\langle  Z(n\beta,n\mu)\rangle\,.$$
The second term in (\ref{4.17}) becomes,

\begin{align}
\begin{split}
      e^{-\mathcal{Z}(x)}= \exp\Bigg\{-\sum_{n=1}^\infty \frac{(-x)^n}{n!}\langle Z^n\rangle_c \Bigg \}\approx \exp\Bigg\{ -\sum_{n=1}^\infty \frac{(-x)^n}{n!}\langle Z(n\beta,n\mu)\rangle_{\alpha,\lambda}  \Bigg\}\,.
\end{split}
\end{align}
\textcolor{black}{In this section, we aim to compute the annealed and quenched free energy. We have all the ingredients to compute the free energy. However, the resurgent analysis of Quenched free energy is a little bit harder, as the genus sum up to infinity cannot be done explicitly in a non-perturbative way. \textit{We perform our computation by taking into account only  $g=0$ and $g=1$ topologies with one asymptotic boundary.}}\\

\begin{figure}
\begin{subfigure}{.50\textwidth}
  \centering
  \includegraphics[width=0.9\linewidth]{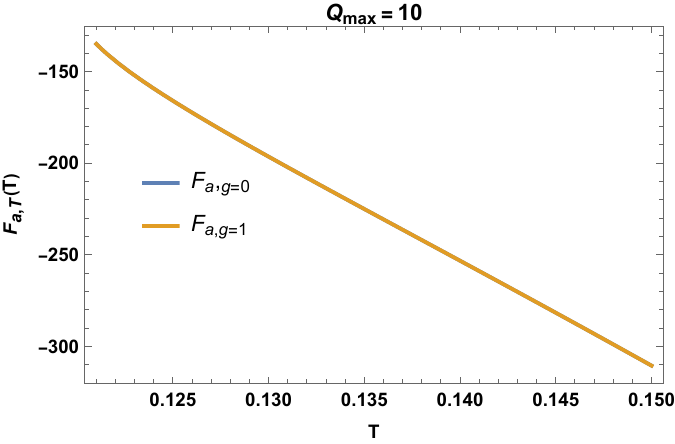}
  \caption{}
  \label{fig:sub2}
\end{subfigure}  %
\begin{subfigure}{0.50\textwidth}
  \centering
  \includegraphics[width=0.9\linewidth]{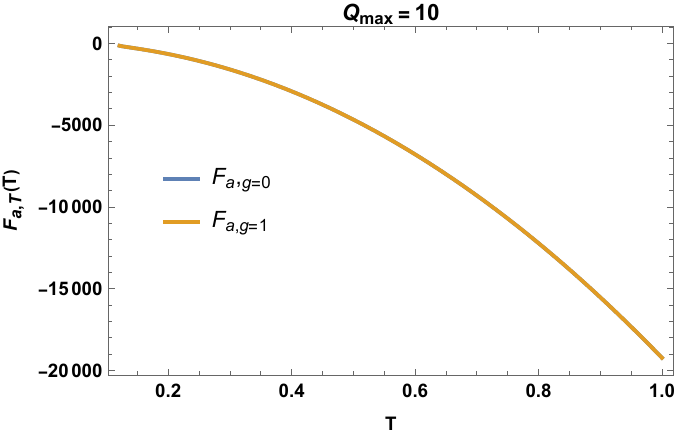}
  \caption{}
  \label{fig:sub42}
\end{subfigure}
\begin{subfigure}{0.50\textwidth}
  \centering
  \includegraphics[width=0.9\linewidth]{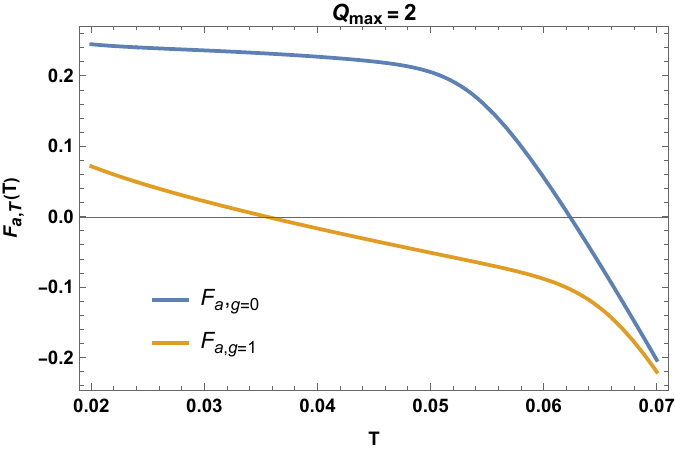}
  \caption{}
  \label{fig:sub43}
\end{subfigure}%
\begin{subfigure}{0.50\textwidth}
  \centering
  \includegraphics[width=0.9\linewidth]{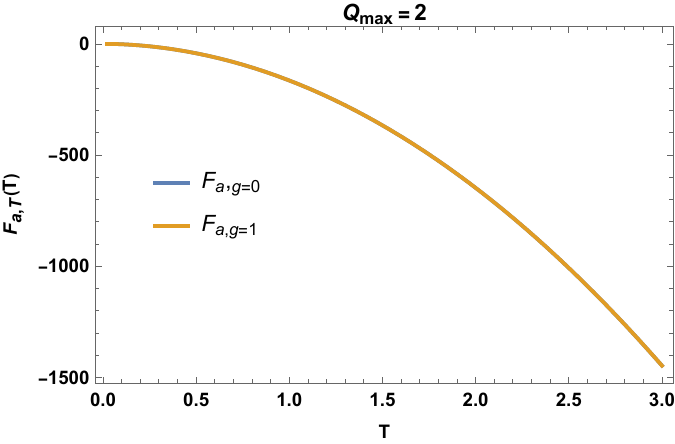}
  \caption{}
  \label{fig:sub44}
\end{subfigure}
\caption{The plot showing the variation of annealed free energy ($F_{a}$) with temperature for four different cases. The value of the parameters chosen are as follows $\alpha =0.003 ,\,\mu =0.05, c=0.2 ,\,\varphi_{b,Q} =Q^3$. Fig.~(\ref{fig:sub2}) and Fig.~(\ref{fig:sub42}) are for $Q_{max}=10$ but for different range of $T\,.$ Similarly, Fig.~(\ref{fig:sub43}) and Fig.~(\ref{fig:sub44}) for $Q_{max}=2$ but for different range of $T\,.$ }
\label{fig2}
\end{figure}

\textbf{Annealed free energy:}
 The `annealed' free energy can be easily computed by taking the logarithm of the one-point function. We compute $F_{a}$ for both genus zero case and genus one case. We make a comparative study of annealed free energy for the deformed one at $g=0$ and $g=1$ in Fig.~(\ref{fig2}). The expressions for the $F_{a}(T)$ has the form:

\vspace{-0.5cm}
\begin{align}
\begin{split}
 {F}_{a}\Big|_{g=0}&=-T \,\log\Big\langle {Z}(\beta,\mu)\Big\rangle_{\alpha,\lambda}\,.
 \\&=-T{\Bigg[ \log\sum_Q \scriptscriptstyle{\left(\frac{\varphi_{b,Q} ^{3/2}e^{-\Delta \Tilde{M_0}(Q)} \exp \left(\frac{2 \pi ^2 \varphi_{b,Q} }{\Delta (T)}\right) \exp \left(\frac{\mu  \left(\Delta (T)-\frac{1}{T}\right)}{\alpha  \Delta (T)}\right)}{\left(\Delta (T)^{5/2}(T \right) (|2 \alpha  Q \Delta (T)-1|)}+\frac{\varphi_{b,Q} ^{3/2} e^{-\xi \Tilde{M_0}(Q)}\exp \left(\frac{2 \pi ^2 \varphi_{b,Q} }{\xi (T)}\right) \exp \left(\frac{\mu  \left(\xi (T)-\frac{1}{T}\right)}{\alpha  \xi (T)}\right)}{\left(\xi (T)^{5/2} \left(T \right)\right) (|2 \alpha  Q \xi (T)-1)|}\right)}}\\& \times e^{S_0(Q)}\,U''(\varphi_0(Q))^{1/4}\Bigg].
\end{split}
\end{align}
%\vspace{0.8cm}
%\newpage
Annealed free energy by taking into account the contributions from higher genus has been computed in Appendix (\ref{B.2}).\\
\begin{figure}[b!]
\begin{subfigure}{0.50\textwidth}
  \centering
  \includegraphics[width=0.9\linewidth]{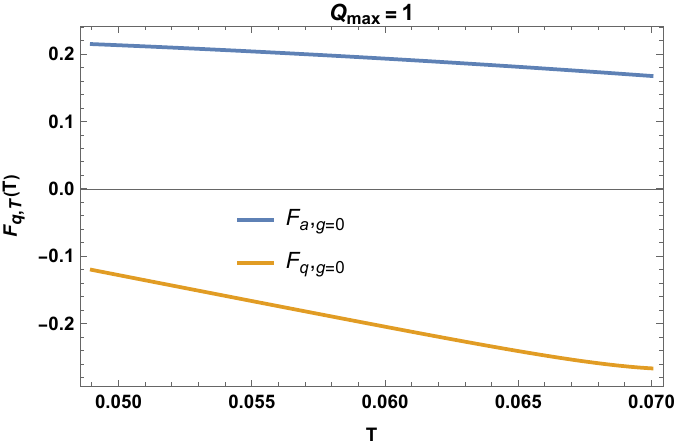}
  \caption{}
  \label{fig:sub31}
\end{subfigure}%
\centering
\begin{subfigure}{0.50\textwidth}
  \centering
  \includegraphics[width=0.9\linewidth]{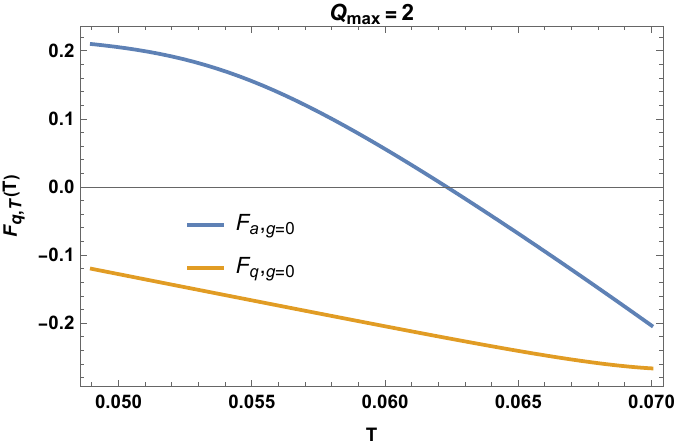}
  \caption{}
  \label{fig:sub32}
\end{subfigure}
\begin{subfigure}{0.50\textwidth}
  \centering
  \includegraphics[width=0.9\linewidth]{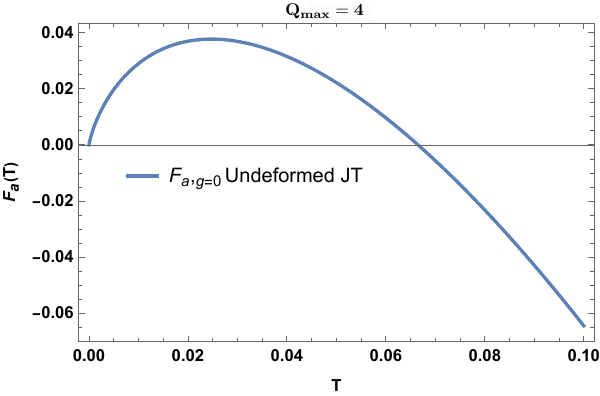}
  \caption{}
  \label{fig:sub332}
\end{subfigure}
\caption{The plot showing the variation of annealed free energy ($F_{a}$)/quenched free energy for the deformed JT-gravity with temperature for different temperature ranges. Fig.~(\ref{fig:sub31}),  Fig.~(\ref{fig:sub32}) and Fig.~(\ref{fig:sub332}) are for  $Q_{max}=1\,,2$  and $4\,.$ The values of the other parameters are chosen as follows: $\alpha =0.01; \mu =0.0005; s=4$ and $\varphi_{b,Q} = Q^3\,.$}\label{fig3}
\end{figure}

$\textbf{Quenched free energy:}$ The calculation of Quenched free energy is more involved than the other because of the sum over $n$ that runs from $1$ to $\infty$ in the low-temperature limit. In low-temperature, the annealed free energy for the deformed JT gravity has a larger gradient than that of the undeformed one.
The computation of quenched free energy ($F_{q}$) has been done using the integral representation of the quenched free energy discussed in (\ref{4.17}) in ``\textit{Mathematica}''. \par

We observe the following: 
\begin{itemize}
    \item From Fig.~(\ref{fig:sub31}) and (\ref{fig:sub32}), it is evident that for $g=0$, the quenched free energy is monotonic, unlike the annealed one. The reason behind the monotonicity is that in the deformed case, we put the lower cutoff of the temperature $T$ (say $T_{c}$) coming from the reality condition of the deformed partition function in (\ref{7.4e}). In general, the non-monotonic behavior only arises at very low temperatures. So, if the non-monotonic behaviour of the quenched free energy in undeformed JT gravity arises at  $T_{n}$ and the cutoff temperature $T_c>T_{n}$ then $F_{q}$ becomes monotonic for $T>T_c$. 
    
    \item However, for  annealed free energy, we observe that for the smaller charge cutoff ($Q_{\textrm{max}}=2$), $F_a$ is monotonic at lower temperature as shown in  Fig.~(\ref{fig:sub43}) when $T_{c}<T_{n}$. We also observe that the monotonicity retains when the charge cutoff becomes large ($Q_{\textrm{max}}=10$) as shown in Fig.~(\ref{fig:sub2}). 
    \item \textit{We further observe that  unlike JT gravity, the contribution from $g=0$ is  sufficient to make the quenched free energy monotonous.} This is shown in Fig.~(\ref{fig:sub332}). 
    
   % \item Last but not the least, at small temperatures and for $g=(0,1)$, the annealed free energy first increases and then monotonically decreases, but the quenched free energy monotonically decreases at a  cut-off of $Q$. However, for $g=(0,1)$, we observe that the monotonicity in annealed free energy gets restored larger $Q_{\textrm{max}}$ like the quenched one.
\end{itemize}

\section{Spectral form factor}\label{sec8}

{Finally, we focus on calculating the spectral form factor (SFF), which will help us analyze the spectral properties of our model. Originally introduced as a probe for analyzing the spectrum of quantum systems, it has developed into an indispensable diagnostics of quantum chaos \cite{Cotler:2016fpe}. It interpolates between early-time probes, such as the out-of-time order correlator (OTOC), and a more standard random matrix-based measure. We compute the SFF both from the disconnected and the connected two-point functions. Like JT gravity, the transition from the ramp to the plateau structure is not clearly visible in our case also without \textit{tau-scaling} \cite{Okuyama:2023pio}. This is a distinctive feature of JT gravity or $T\Bar{T}+J\Bar{T}$ deformed 2D gravity, unlike random matrix models. Previously, there have been efforts to show the ramp to the plateau transition by the inclusion of branes \cite{Castro:2023rfd}. \textit{However, recently by the method of $\tau$ scaling \cite{Okuyama:2023pio}, the possible transition to the plateau is visible in JT gravity, and in our case, also.} The contribution to the ramp comes from the wormhole contribution again.}
%\vspace{0.4 cm}
\begin{figure}[t!]
\begin{subfigure}{0.50\textwidth}
  \centering
  \includegraphics[width=0.9\linewidth]{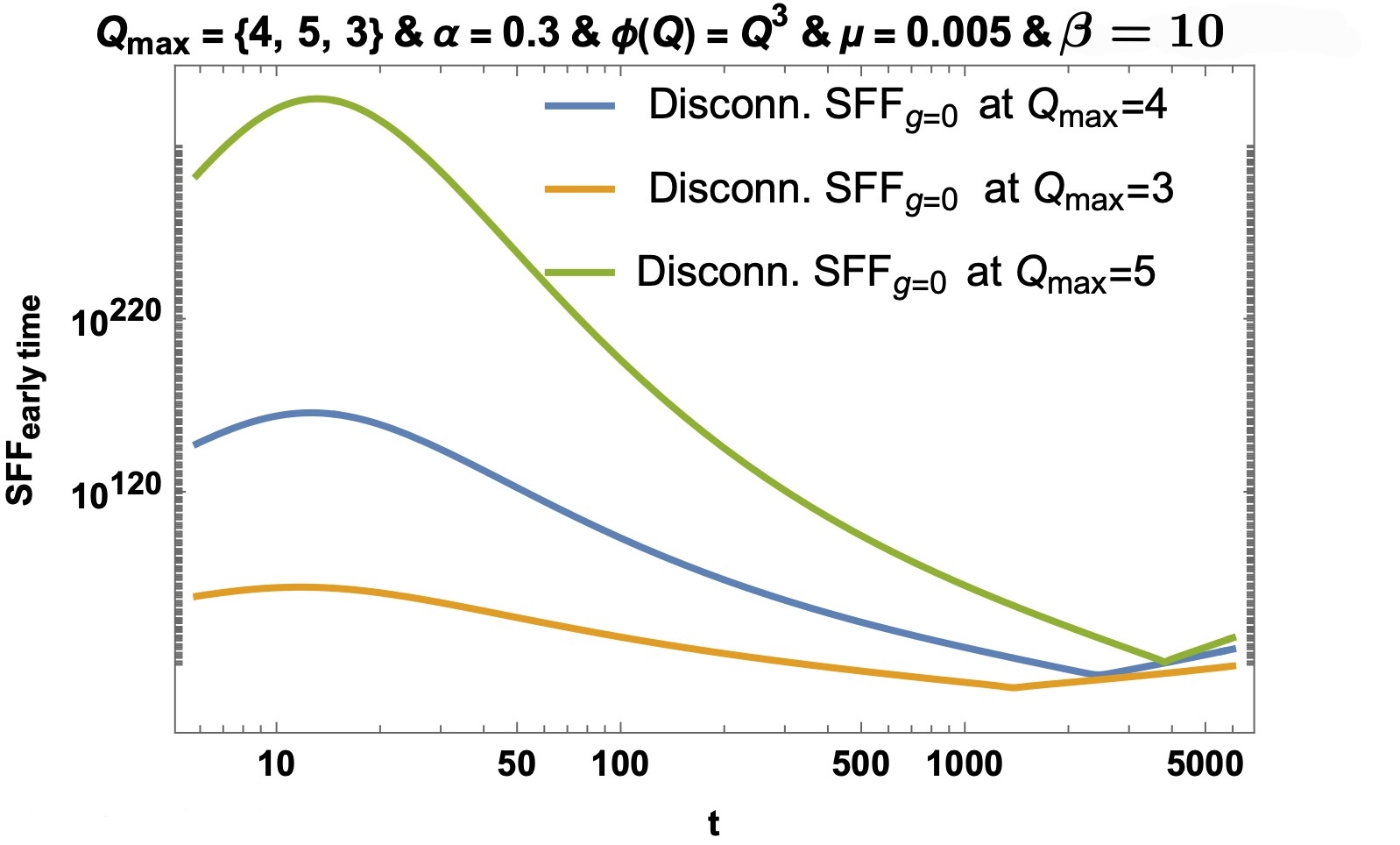}
  \caption{}
  \label{fig:sub5a}
\end{subfigure}%
\centering
\begin{subfigure}{0.45\textwidth}
  \centering
  \includegraphics[width=0.9\linewidth]{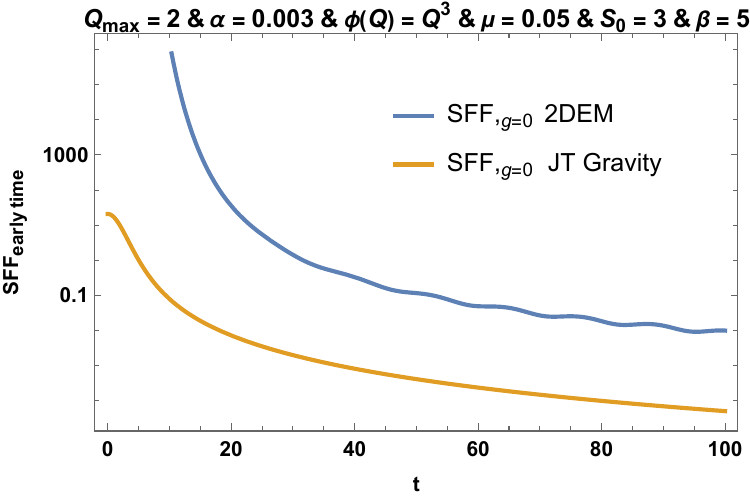}
  \caption{}
  \label{fig:sub5c}
\end{subfigure}
\begin{subfigure}{0.5\textwidth}
  \centering
  \includegraphics[width=0.9\linewidth]{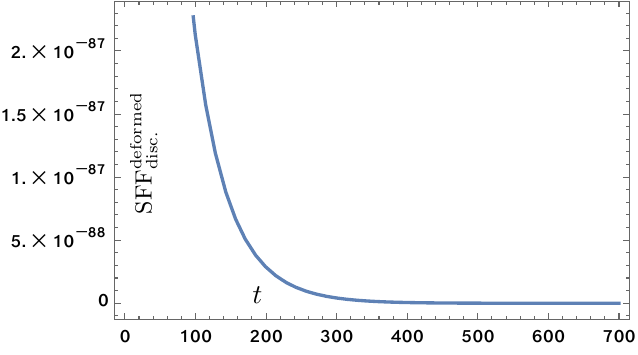}
  \caption{}
  \label{fig:sub5l}
\end{subfigure}
\caption{Plot showing the variation of disconnected SFF as a function of time. Fig.~(\ref{fig:sub5a}) depicts that increasing the sum up to large value of charge will make the  disconnected SFF decay up to larger times like JT gravity (Fig.~(\ref{fig:sub5c})). Plot (\ref{fig:sub5l}), showing 2DEM deformed disconnected SFF for the good sign of defromation parameter, $\alpha>0.$  } \label{fi4nw}
\end{figure}
\\\\
\textcolor{black}{\textbf{Comments on the time scale of decaying nature of disconnected SFF in the deformed theory:}
 It has been argued in \cite{Anegawa:2023klh,Saad:2019pqd} that 
the factorization of SFF, $\langle|Z(\beta+it)|^2\rangle\approx\langle Z(\beta+it)\rangle \langle Z (\beta- it)\rangle$ is valid within a certain time scale, $t\ll e^{S_{0}/2}$ and up to this time scale the SFF has has a decaying nature. The total SFF is well approximated by the disconnected SFF only upto $t\ll e^{S_{0}/2}.$ Our computation shows that this is indeed true for deformed JT-gravity also. \textit{From Fig.~(\ref{fig:sub5a}), it is evident that the deformed disconnected SFF has a decaying nature up to the timescale mentioned above $(i.e. \,\, e^{S_{0}/2})$. We also notice that the disconnected undeformed 2DEM SFF, as depicted in Fig.~(\ref{fig:sub5c}), decays like pure JT gravity. Furthermore, if we make the charge sum larger for the deformed theory, we find that the transition point of the disconnected SFF from decay to growth is pushed to a much later time, as shown in Fig.~(\ref{fig:sub5a}). Eventually, we expect that if one can manage to perform a sum over charge up to $\infty$ for $\alpha>0$ (small values), the disconnected SFF ceases growing. We also show that for $\alpha < 0$ the disconnected SFF never grows as shown in (\ref{fig:sub5l}). }}

%\begin{figure}[htb!]
%\centering
  %\includegraphics[width=0.7\linewidth]{q_5_4_3.pdf}
%\caption{Plot showing the variation of disconnected SFF as a function of time. The picture depicts that increasing the sum up to large value of charge will make the disconnected SFF decay up to larger times.  } 
%\label{fig888}
%\end{figure}

%\newpage
\textbf{Connected SFF:} We also calculate the connected two-point spectral form factor in the tau-scaling limit. The connected SFF is given by \begin{align}
    \langle Z(\beta+it,\mu)Z(\beta- it,\mu)\rangle\,.
\end{align}

\begin{figure}[htb!]
\centering
\begin{subfigure}{0.50\textwidth}
  \includegraphics[width=0.9\linewidth]{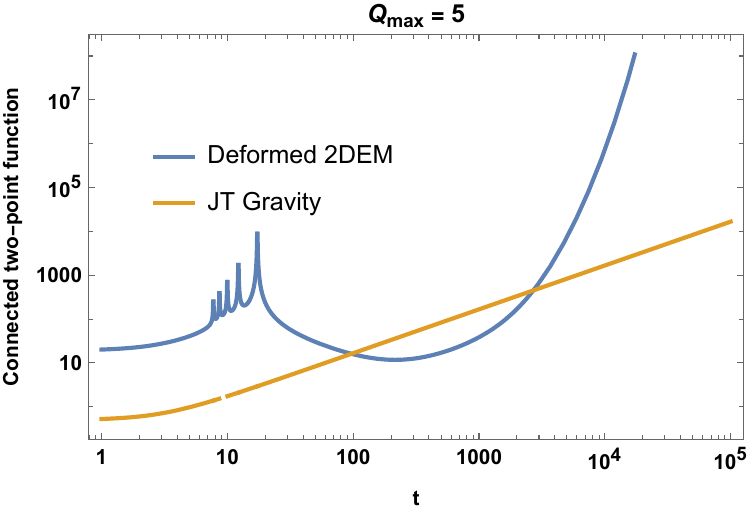}
  \caption{}
  \label{fig:sub52}
\end{subfigure}%
\centering
\begin{subfigure}{0.50\textwidth}
  \centering
 \includegraphics[width=0.9\linewidth]{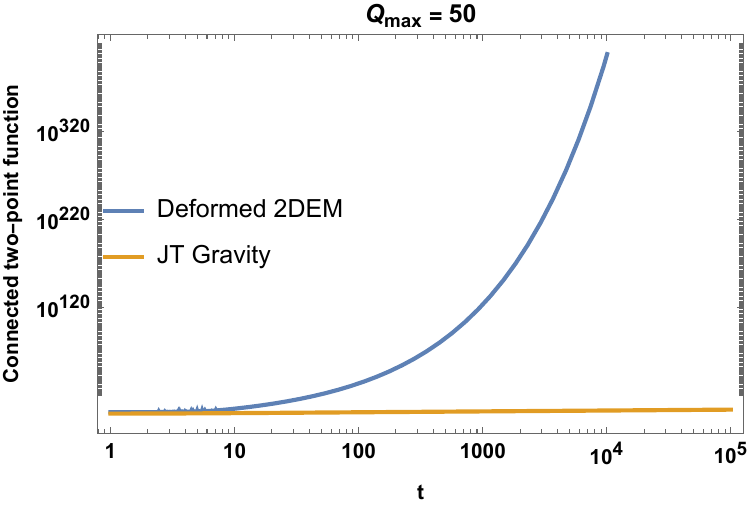}
  \caption{}
  \label{fig:sub4}
\end{subfigure}%
\caption{We have explicitly shown the LogLogplot of Connected two-point function structure for JT gravity and $T\bar{T} +J\bar{T}$ deformed 2D gravity. The maximum value of the charge taken $Q_{max}=5$  and $Q_{max}=50$ in Fig.~(\ref{fig:sub52}) and Fig.~(\ref{fig:sub4}). The other parameters are taken as follows. $\alpha=0.01,\beta =3; \mu =0.005, c=0.002$. }\label{fignnnew5}
\end{figure}

\begin{figure}[b!]
\centering
\begin{subfigure}{0.50\textwidth}
  \includegraphics[width=0.9\linewidth]{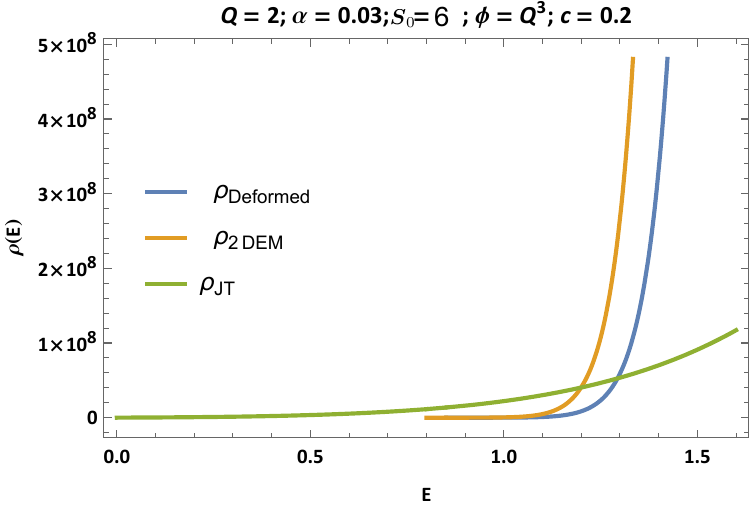}
  \caption{}
  \label{fig:sub3x}
\end{subfigure}%
\begin{subfigure}{0.50\textwidth}
  \includegraphics[width=0.9\linewidth]{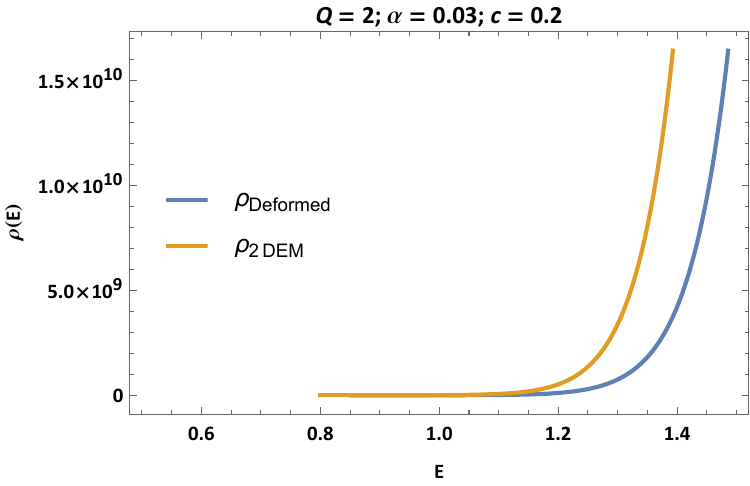}
  \caption{}
  \label{fig:sub3xy}
\end{subfigure}
\caption{Plot showing the variation of density of states as a function of energy. We have chosen the following values of parameters. $\alpha =0.03; S_{0}=6; Q=2$;c=0.2;$\varphi_{b,Q} =Q^3$. Due to deformation and non-zero deformation parameters, it shows an energy gap, unlike undeformed theory. Fig.~(\ref{fig:sub3x}) shows the density of states from both the deformed and undeformed (JT) theory. Fig.~(\ref{fig:sub3xy}) shows that the density of state for both the 2DEM and deformed 2DEM doesnot vanish as $E\rightarrow 0\,.$  }  \label{fig5nw}
\end{figure}

We plot this in Fig.~(\ref{fignnnew5}). In tau-scaling, we just zoom robustly into the larger and larger time scale. We write $\tau=t\exp(-S_0)$ and take $t\rightarrow \infty$ and $exp(-S_0)\equiv \hbar\rightarrow 0$ keeping $\tau$ finite. In general, as stated earlier, the perturbative genus sum upto $\infty$ cannot be done explicitly.
This scaling makes the disk contribution dominant, making the genus sum over other topologies very sub-leading \cite{Blommaert:2022lbh}. Finally, it helps the SFF to reach the \textit{plateau} at a late time. Below we give details of our computation for the the tau-scaled spectral form factor, which takes the following form, 
\begin{align}
Z(\beta+i t,\beta-i t)_{conn}=\frac{e^{S_0}}{4\pi\beta}\int_0^\tau dfe^{-\beta E(f) }
\end{align}
with $f\equiv 2\pi e^{-S_0}\rho_0(E)$.
For JT gravity, the density of the state  is $\rho_0(E)
=e^{S_0}\frac{\sinh(2\pi \sqrt{E})}{4\pi^2}$ (shown in Fig.~(\ref{fig5nw}).). So, the inverse spectrum is given by :
$$E(f)=\frac{\text{arcsinh}^2(2\pi f)}{4\pi^2}\,.$$

Now for our case, we do the inverse Laplace transform the partition function to find $$\rho_0(E,Q)|_{\textrm{2DEM}}=e^{S_0(Q)}\,U''(\varphi_0(Q))^{1/4}\frac{\varphi_{b,Q}  \sinh \left(2 \sqrt{2} \pi  \sqrt{(E-\Tilde{M_0}(Q)) \varphi_{b,Q} }\right)}{\sqrt{2} \pi ^{3/2}}\,.$$

Now, for the $T\bar{T} + J\bar{T}$ deformed case, the energy spectrum in the leading order is given by,
\vspace{-0.9cm}

\begin{align}
    \begin{split}
       \mathcal{E} (\alpha,\lambda)=-\frac{2}{2\lambda-\alpha^2}\Big[1-\alpha Q-\sqrt{(1-\alpha Q)^2+E (2\lambda-\alpha^2)}\Big]\,.
    \end{split}
\end{align}

Now, we proceed to calculate the deformed density of state for the {2D} gravity coupled to the U(1) gauge field.
%\vspace{-0.3cm}
We focus on the spectral density that comes from the disk contribution of the partition function. The density of states is defined as follows,
\begin{align}
    \begin{split}
        \langle Z(\beta,\mu) \rangle_{0}=\int dE\, \rho_{0}(E)e^{-\beta E+\beta \mu Q}\,.
    \end{split}
\end{align}
Now our goal is to find the deformed density of states $\rho_{d}(E,\mu)$. For that, one can write the deformed partition function as follows,
\begin{align}
    \begin{split}
        \langle Z(\beta,\mu)\rangle_{d} &= \int d{E}\, \rho_{0}({E} ) e^{-\beta \mathcal{E}(E)+\beta \mu Q}\,,\\ &
        =\int d\mathcal{E} \,\frac{d{E}} {d\mathcal{E}} \rho_{0}(E(\mathcal{E})) e^{-\beta \mathcal{E}+\beta \mu Q}\,.
    \end{split}
\end{align}
Hence, the deformed density of states in the limit $h\rightarrow \infty \,(\text{i.e } 2\lambda\rightarrow \alpha^2)$ can be written as,
\begin{align}
\begin{split}
    \rho_{d}(\mathcal{E})&=\frac{dE}{d\mathcal{E}}\rho_{0}(E(\mathcal{E}))\,,\\ &
    =(1-\alpha Q) \rho_{0}[\mathcal{E}-\alpha Q \mathcal{E}]\,,\\ &
    =(1-\alpha Q) \,U''(\varphi_0(Q))^{1/4}\frac{e^{S_{0}(Q)}\varphi_{b,Q}}{\sqrt{2}\pi^{3/2}} \sinh\Bigg\{2\sqrt{2}\pi \sqrt{(\mathcal{E}-\alpha Q \mathcal{E}-\Tilde{M_0}(Q))\varphi_{b,Q}}\Bigg\}\,.\label{8.6}
    \end{split}
\end{align}
Here, in (\ref{8.6}) $\mathcal{E}$ is a dummy index and can be replaced by $E$. The inverse spectrum can be written as,
\begin{align}
\begin{split}
E(f)=\frac{\text{arcsinh}^{2}\left(\frac{f \, U''(Q)^{-1/4} \sqrt{\pi}  }{\sqrt{2}\varphi_{b,Q}(1-\alpha  Q)}\right)}{8\pi ^2 \varphi_{b,Q}(1-\alpha Q)}+\frac{\Tilde{M_0}(Q)}{1-\alpha Q}\,.
\end{split}
\end{align}
Now for this $T\bar{T}+J\bar{T}$ deformed spectrum we have,
\vspace{-0.3cm}
\begin{align}
Z(\beta+i t,\beta-i t)_{\textrm{conn}}=\sum_Q\frac{e^{S_0(Q)+2\beta\mu Q}}{4\pi\beta}\int_0^\tau df \, \exp \Bigg(-2\beta \Bigg[\frac{\text{arcsinh}^{2}\left(\frac{f \,  U''(Q)^{-1/4}\sqrt{\pi}  }{\sqrt{2}\varphi_{b,Q}(1-\alpha  Q)}\right)}{8\pi ^2 \varphi_{b,Q}(1-\alpha Q)}+\frac{\Tilde{M_0}(Q)}{1-\alpha Q}\Bigg]\Bigg)\,.
\end{align}

Defining $w=\sqrt{E(f)}$, we have

%\vspace{-0.73cm}

\begin{align}
    \begin{split}
    &   \langle Z(\beta+i t)Z(\beta-i t)\rangle_{conn}=\sum_{Q}\frac{e^{S_0(Q)+2\beta \mu Q}}{4\pi^{3/2}\beta }\\ &
\int_{\lambda_1}^{\lambda_2}dw\,{{\frac{w \exp \left(-2 \beta  w^2\right) \left(4\sqrt{\pi}U''(\varphi_0)^{1/4}\varphi_{b,Q} ^{3/2} (1-\alpha  Q)^{3/2}\right) \cosh \left(2\pi\sqrt{2\varphi_{b,Q}((1-\alpha Q)w^2-\tilde M_0(Q))}\right)}{\sqrt{w^2-\frac{\tilde M_0(Q)}{1-\alpha  Q}}}}\,},\\&
= \sum_{Q}\frac{\sqrt{2} \sqrt[4]{3}\, Q^{7/2} (\alpha  Q-1)^{3/2} e^{\frac{2 \beta \tilde{M}_0(Q)} {\alpha  Q-1}+\frac{\pi ^2 Q^3 (1-\alpha  Q)}{\beta }} \exp \left(Q^2+2 \beta  \mu  Q\right)}{8\pi  \beta^{3/2}}[\text{erf}(\sigma_{+})+\text{erf}(\sigma_{-})]\,,
    \end{split}
\end{align}
with,
\begin{eqnarray}
  \lambda_1=\sqrt{\frac{\tilde M_0(Q)}{1-\alpha Q }}   \,\,\,,\lambda_{2}= \sqrt{\frac{\text{arcsinh} ^{2}\left(\frac{ \sqrt{\pi } u''(\varphi_0)^{-1/4} \tau }{\sqrt{2} \varphi_{b,Q}  (1-\alpha  Q)}\right)}{8 \pi ^2 \varphi_{b,Q}  (1-\alpha  Q)}+\frac{\tilde M_0(Q)}{1-\alpha  Q}}
\end{eqnarray}
and,
\begin{align}
\sigma_{\pm}=\frac{\pm 2 \pi ^2 Q^3 (\alpha  Q-1)+\beta  \sqrt{\text{csch}^{-1}\left(\frac{2 \sqrt{2} \sqrt[4]{3} Q^2 (\alpha  Q-1)}{\tau }\right)^2}}{2 \pi  \sqrt{\beta } Q^{3/2} \sqrt{1-\alpha  Q}}.
\end{align}

We have plotted it in Fig.~(\ref{fig 8}), which shows a convergent plateau structure in leading order.\\\\

\begin{figure}[htb!]
\begin{subfigure}{0.50\textwidth}
  \centering
  \includegraphics[width=0.9\linewidth]{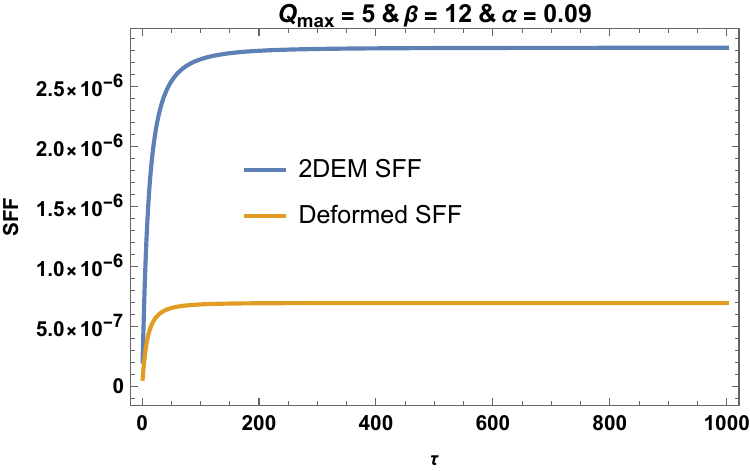}
  \caption{}
  \label{Fig:77a}
\end{subfigure}%
\centering
\begin{subfigure}{0.50\textwidth}
  \centering
  \includegraphics[width=0.9\linewidth]{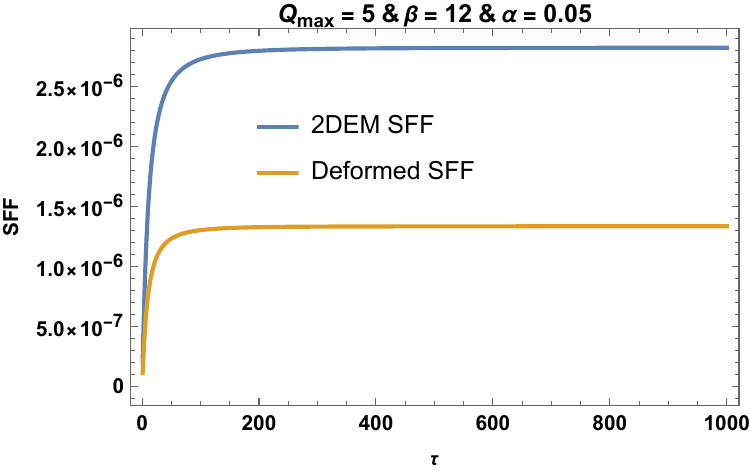}
  \caption{}
    \label{Fig:77b}
\end{subfigure}%
\caption{The plot shows the plateau structure at $g=0$ for $T\Bar{T}+J\Bar{T}$ deformed JT gravity. The plateau structure remains constant for large $\tau=t \exp(S_0)$. Value of other parameters are as follows, $\mu=0.05, c=0.5, \varphi_{b,Q}=Q^3 .$ Fig.~(\ref{Fig:77a}) and Fig.~(\ref{Fig:77b}) are for two different values of $\alpha\,.$
 }
\label{fig 8}
\end{figure}

%\newpage

{\bf Comments on Late time topology and physicality of SFF:}
The SFF has a representation in terms of the \textit{TFD (Thermofield Double States) }states \cite{Blommaert:2018iqz}. One can relate it to the \textit{return amplitude} in the following way,
\begin{align}
\begin{split}
\textrm{Tr}[e^{-(\beta+it)H}]=\langle TFD_{\beta}|e^{-\frac{it}{2}(H_{L}+H_{R})}| TFD_{\beta}\rangle\,.
\end{split}
\end{align}
The spectral form factor is simply the return amplitude to the TFD. We therefore analytically continue the renormalized boundary length to $\beta + it $ and gluing it back to Euclidean renormalized boundary of total length  $\beta$ as shown in Fig.~(\ref{fig9}). Now we have 
$$|Z(\beta+it)|^2_{2DEM}\approx Z(\beta+it)_{2DEM}Z(\beta-it)_{2DEM}\,\,\,\,\,\,\,\,\,\,\,\, \text{ for } t \ll{e^{\frac{S_0}{2}}}\,.$$

\begin{figure}[h!]
    \centering
  \includegraphics[scale=0.18]{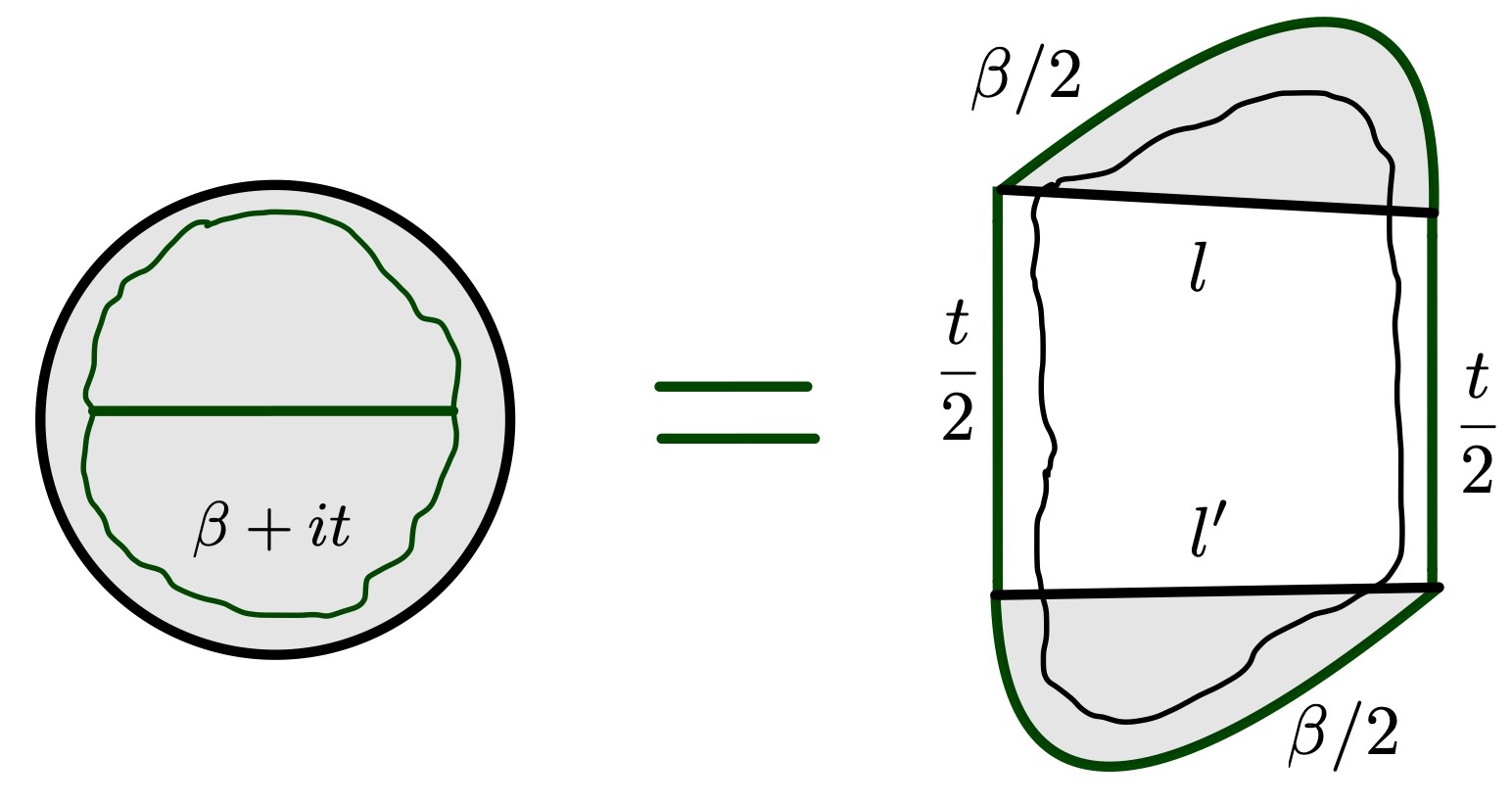}
    \caption{Complexified hyperbolic space with disk topology and boundary of length $\beta+it$. The right side geometry is half Euclidean and half Lorentzian. The half disks prepare the Hartle hawking state. $l$ and $l'$ are renormalized lengths.}
    \label{fig9}
\end{figure}

This description of factorization is only valid for early time scales. The wormhole or double-trumpet topology connects two points in spacetime that are far away via a short geodesic. As we have a two-sided AdS black hole in the background, there is a dual TFD state, which is also dual to the \textit{Hartle-Hawking state}. \textcolor{black}{Initially, the Hartle-hawking state has support at a certain length of the \textit{Einstein-Rosen Bridge(ERB)}.} As time evolves, the JT universe length shrinks and emits a `baby universe' (A third quantized description) \cite{Saad:2019pqd} which may be reabsorbed at some late time or may not be reabsorbed at late time.\par
The reabsorption phenomena do not lead to the decay of the length of the ERB, but in case it emits a closed baby universe, the ERB length decreases. This leads to a dominant contribution in calculating the amplitude of \textit{return probability} or SFF. This physically implies the ramp structure of the total SFF. But in early times, the ERB grows. After calculating the return probability, we see it does not lead to giving the dominant contribution as the initial state contains a small ERB, and the final state contains a large ERB without the emission of a closed baby universe and that suggests that the probability of emission of the baby universe (for initial time scale) is small.\par

\begin{figure}[htb!]
    \centering
   \includegraphics[scale=0.17]{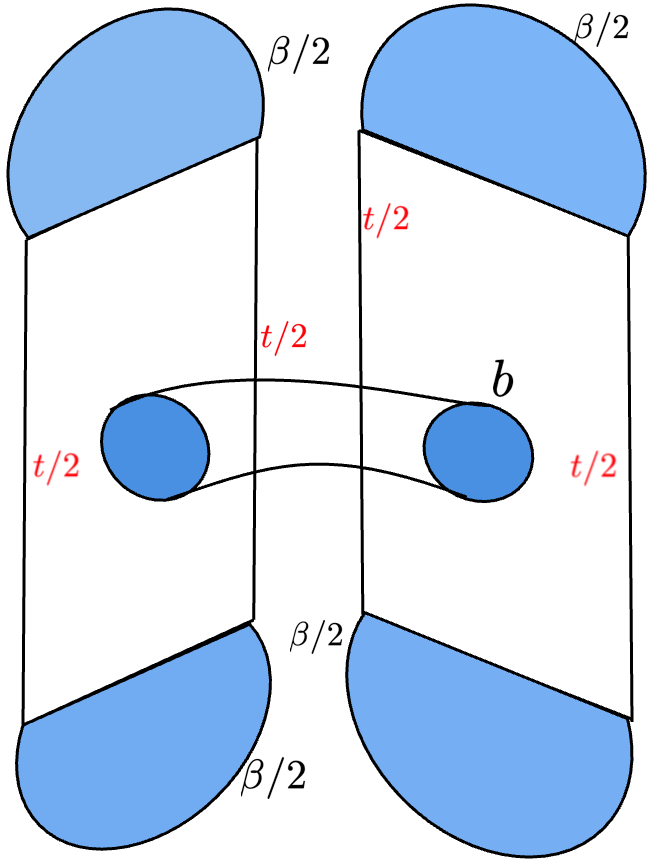}
    \caption{Schematic figure describing the trading of a single baby universe of length $b$. The half discs are of length $\beta/2$ and in between there is a Lorentzian time evolution for time $t$. }
    \label{fig11}
\end{figure}

We leave a detailed investigation to make this more precise for future work. We need to calculate the overlap of the time-evolved Hartle-Hawking state in the length basis ($\psi_E^D(l')$). %One needs to vary the total action and calculate the boundary stress tensor. The Hamiltonian used for the time evolution is the $`00$' component of the boundary stress tensor. 
Then, one needs to solve the Schrödinger equation to find the time-evolved Hartle-Hawking state for  JT gravity coupled to U(1) gauge theory. As mentioned before, Hartle-Hawking states will have a support at ERB, the length of which is small initially. Finally, we need to compute $\langle l,b|l' \rangle $, as defined below, to find the probability of emission of a baby universe of length $b$  as in Fig.~(\ref{fig11}). 
%or that purpose, we need to know the deformed Hartle-Hawking state in the length basis.
$$\langle l,b|l' \rangle = \int_0^\infty dE\, \xi(E) \, \psi_E^D(l)\psi_E^D(l')\,, $$
where %$\psi_E^D(l)$ is the deformed Hartle-Hawking state of the 2DEM theory in the length basis and 
$\xi(E)$ has to be determined from the boundary-particle formalism in AdS at finite cutoff \cite{Yang:2018gdb}.\par 
\par  \textit{Now we make the following interesting observation. For our case, the ramp of the SFF grows faster than the undeformed JT gravity, which means the possibility of the emission of a single baby universe becomes large (for late times) as we couple the theory of JT gravity with the $U(1)$ gauge group non-minimally. As suggested in \cite{Saad:2019pqd}, the minimal coupling of the matter fields does not lead to any backreacted geometry, but we expect non-minimal couplings can do so. As the trading of multiple baby universes will lead to the decay of the SFF, the probability of trading multiple baby universes in our case is vanishingly small in later time as expected from the higher rate of growth of the ramp compared to the usual JT gravity.}\par

\section{Discussions and outlook}\label{sec9}

Mainly motivated by recent developments in topological gravity \cite{Saad:2019lba, Iliesiu:2020zld, Iliesiu:2019lfc}, in this paper, we consider the 2D-dilation gravity with nonlinear dilaton potential coupled with $U(1)$ matter field. \textcolor{black}{Eventually, apart from computing the partition functions, we focus on different thermodynamic properties due to the $U(1)$ field.} We started by calculating the partition function of the $U(1)$ coupled theory. Then, we move on to the computation of correlators involving the partition function, and fromthat, we investigate free energy and spectral form factors. After that, we deform the boundary theory by a linear combination of two irrelevant deformations $T\Bar{T}$ and $J\Bar{T}$ and investigate the qualitative changes of those above-mentioned thermodynamic properties. Below, we summarize the main findings of our paper.
\begin{itemize}
\item First, we compute the partition function for $2D$-gravity coupled with $U(1)$ and $SO(3)$ matter field. We also discuss how the particular theory comes from the dimensional reduction of Einstein-Maxwell theory in four dimensions. Except for the usual linear term involving the dilaton field, we have some extra terms that are non-linear in the dilaton. We compute the Zeta-regularised (in dilaton integral) partition function. 
%The saddle point condition sets the Ricci scalar of the manifold to a negative constant depending on the saddle point itself.

\item Second, we calculate the one and two-point function of the partition function of the theory in perturbative topological expansion. We also comment on the main differences between the JT-gravity and the $U(1)$ coupled one that we considered here.
\item \textcolor{black}{Then, we consider the $T\bar{T}+J\Bar{T}$ deformation of the dual boundary theory and then, using the same deformation kernel, we compute the deformed partition function and its one and two-point function for the  $U(1)$ coupled 2D gravity theory.} We notice that the one and two-point functions have divergence because of the hard cutoff of the boundary length $\beta/\epsilon$. We discuss how one can cancel this divergence by changing the boundary condition suitably.
\item Eventually, we compute different thermodynamic quantities such as annealed and quenched free energy. We investigate the quenched free energy at low chemical potential and low temperatures. We also motivated how the $n$-point correlator gets reduced to a one-point correlator at this specific limit. We show that the annealed free energy has monotonic behaviour at low temperatures. Also, the quenched free energy at  $g=0$ is monotonic independent of the upper cutoff of the sum over $U(1)$ charge in the path integral.

\item  \textcolor{black}{We then move to the computation of the Spectral Form Factor (SFF), which encodes the chaotic nature of the theory. We explicitly discuss the ramp-plateau transition of SFF for both the JT and deformed JT gravity. \textit{We show that in early time up to a specified time scale ($t\ll e^{S_{0}/2}$), the deformed disconnected  SFF has a decaying nature like pure JT gravity and the undeformed 2DEM theory}, for which the disconnected SFF asymptotically approaches zero for large time.  Also, for $T\Bar{T}+J\Bar{T}$ deformation we only take the $\mathcal{O}(1/h)^0$ (leading order in $1/h$) term. It would be worth investigating the $1/h$ corrections to the disconnected SFF. We leave this investigation for the future.}
\item  We also compute the connected SFF from the deformed two-point function. Unlike the pure JT gravity, the early time-connected two-point function shows oscillation in the dip region. \textit{We show the ramp of the SFF in the deformed theory increases faster than the undeformed one} and discuss a plausible physical interpretation of it by connecting with the fact that there are more chances of creating baby universes when we couple the theory with $U(1)$ matter field.

\end{itemize}
Now, we end by stating some possible future directions. \textcolor{black}{It is promising and also very interesting to perform the computation of the deformed partition functions purely from the bulk side. However, the computation of the glueing measure and moduli space volume is more non-trivial in the second-order formalism of JT gravity. It requires a detailed study, and we hope to make some progress in that direction in the near future.} 
%\textcolor{black}{First and foremost, we concentrate on doing the path integral in a general setting without going to the saddle point approximation and investigate the qualitative differences in the different thermodynamic and chaotic behaviour of the model. } 
We also wish to compute the matrix model dual to this deformed theory along the line of \cite{Jafferis:2022wez, Maxfield:2020ale} using the technologies mentioned in \cite{Saad:2019lba}.  \textcolor{black}{We have observed the ramp-plateau behaviour of SFF for our deformed theory, and it is an encouraging sign that perhaps a dual matrix model description exists. We hope to report on it in the near future.}  From that, one can compute the SFF, which could be a potential cross-check to our computations. Also, for our case, we have computed the $j=0$ sector of the $SO(3)$ gauge group. One straightforward generalization is to compute the partition function for $j\ne 0$ and then compute the free energy, SFF, etc. One can compute OTOC for this deformed theory by inserting bi-local operators on the disk and check the Lyapunov growth \cite{Mertens:2019tcm}. One can also eventually try to calculate different thermodynamic aspects like free energy and specific heat by the inclusion of FZZT branes (with non-vanishing correlation) on the geodesic boundaries along the line of \cite{Castro:2023rfd, Blommaert:2021fob}. Furthermore, the aspect of the calculation of the Hartle-Hawking wavefunction is also non-trivial, but if it can be computed, one can gradually proceed to calculate the late time complexity for 2D gravity theory coupled to U(1) gauge theory. Also, as discussed in the section~(\ref{sec8}), it will also help us to understand the higher rate of growth of the ramp compared to the usual JT gravity by computing the probability of emission of the baby universe.  Apart from these, we also wish to inspect the \textit{Eigenstate thermalization hypothesis} (ETH) \cite{Jafferis:2022wez}, from the inverse Laplace transform of matter correlators on the boundary of $AdS_2$, to determine the non-diagonal elements of the dual matter operators in the energy-basis. Last but not least, it might be interesting to repeat our analysis for hyperscaling violating geometry \cite{Khoeini-Moghaddam:2020ymm}.
%\vspace{-0.50  cm}
%\newpage
\section*{Acknowledgements}
  We thank Thomas G. Mertens for useful email correspondence. We also thank Ashoke Sen for insightful discussions on the addition of counter terms. S.G and S.P would like to thank Poulami Nandi for useful discussions and pointing out the reference \cite{Hatfield:1992rz}. A.B would like to thank the speakers and participants of the workshop ``Quantum Information in QFT and AdS/CFT-III" organized at IIT Hyderabad between 16-18th September, 2022 and funded by SERB through a Seminar Symposia (SSY) grant (SSY/2022/000446) and  ``Quantum Information Theory in Quantum Field Theory and Cosmology" between 4-9th June, 2023 hosted by Banff International Research Centre at Canada for useful discussions. Research of S.G is supported by the Prime Minister's Research Fellowship (ID:1702711) by the Ministry of Education, Government of India. A.B is supported by Relevant Research Project grant (202011BRE03RP06633-BRNS) by the Board Of Research In Nuclear Sciences (BRNS), Department of Atomic Energy (DAE), India, Core Research Grant (CRG/2023/001120) and Mathematical Research Impact Centric Support Grant (MTR/2021/000490) by the Department of Science and Technology Science and Engineering Research Board (India). A.B would like to thank the FISPAC Research Group, Department of Physics, University of Murcia, especially, Jose J. Fernández-Melgarejo, for hospitality during the course of this work. A.B. also acknowledge the associateship program of the Indian Academy of Science, Bengaluru.
%  \newpage
\appendix
\section{Details of the calculation in section \ref{section 3} }\label{a1}
\textbf{$\frac{1}{h}$ correction to the deformed disk partition function:} Similarly, extending (\ref{6.6}) the $\mathcal{O}(\frac{1}{h})$ correction is given by:

\begin{align}
\begin{split}
{\mathcal{I}^{disk}_{\frac{1}{h}}}
&=\int_0^{\infty}d\beta'\exp\Bigg[ 
  \frac{(\beta'-\beta)\mu}{\alpha \beta'} \Bigg]\frac{\beta}{8\alpha^3\beta'^3 h}\Bigg\{\mu ^2 \delta (t)-2 \mu  \delta '(t)+\delta ''(t)\Bigg\}\Bigg|_{t=\frac{\beta-\beta'}{\alpha \beta'}+\beta'Q} \\ & \hspace{3 cm}\times\frac{\varphi_{b,Q}^{3/2}}{\sqrt{2\pi}{\beta'}^{3/2}}\, e^{\frac{2\pi^2\varphi_{b,Q}}{\beta'}+\beta' \mu' Q}\,,\\&
 =\mu^2\frac{\beta}{8\alpha^2}\Bigg[\int_0^\infty d\beta'\frac{1}{\beta'^3}\exp{\frac{(\beta'-\beta)\mu}{\alpha \beta'}}\Bigg\{ \frac{\delta(\beta'-\Delta)}{|f'(\Delta)|}+\frac{\delta(\beta'-\xi)}{|f'(\xi)|}\Bigg\} \frac{\varphi_{b,Q}^{3/2}}{\sqrt{2\pi}{\beta'}^{3/2}}\, e^{\frac{2\pi^2\varphi_{b,Q}}{\beta'}+\beta' \mu' Q}\Bigg]\\&
 +\Bigg[\frac{2\mu\beta}{8\alpha^2 h}\int_0^\infty d\beta'\frac{\exp{\frac{\mu  \left(\beta '-\beta \right)}{\alpha  \beta '}} \left(\beta  \mu -3 \alpha  \beta '\right)}{\alpha  Q \left(\beta '\right)^4-\beta  \left(\beta '\right)^2}\Big\{\frac{\delta(\beta'-\Delta)}{|f'(\Delta)|}+\frac{\delta(\beta'-\xi)}{|f'(\xi)|}\Big\}\Bigg] \times\frac{\varphi_{b,Q}^{3/2}}{\sqrt{2\pi}{\beta'}^{3/2}}\, e^{\frac{2\pi^2\varphi_{b,Q}}{\beta'}+\beta' \mu' Q}\\ &+\Bigg[\frac{\mu ^2 e^{\frac{\mu  \left(\beta '-\beta \right)}{\alpha \beta '}}}{\left(\alpha \beta '\right) \left(\beta '\right)^3}-\frac{6 \mu  e^{\frac{\mu  \left(\beta '-\beta \right)}{\alpha \beta '}}}{\left(\beta '\right)^4}+\frac{12 \alpha e^{\frac{\mu  \left(\beta '-\beta \right)}{\alpha \beta '}}}{\left(\beta '\right)^4}\Bigg]\Bigg\{ \frac{\delta(\beta'-\Delta)}{|f'(\Delta)|}+\frac{\delta(\beta'-\xi)}{|f'(\xi)|}\Bigg\}\frac{\varphi_{b,Q}^{3/2}}{\sqrt{2\pi}{\beta'}^{3/2}}\, e^{\frac{2\pi^2\varphi_{b,Q}}{\beta'}+\beta' \mu' Q}\,.
 \end{split}
 \end{align}

Then, 
 \begin{align} Z^{disk}_{\textrm{2DEMSO(3)}}|_{\frac{1}{h}}=\mathcal{I}^{disk}_{\frac{1}{h}} \times (\text{Other U dependent factors}) \,.
 \end{align}

 Similarly, we can calculate the trumpet partition function and eventually calculate the one and two point functions with $\frac{1}{h}$ corrections. Now, doing the path integral using the steepest descent method over the dilaton field and the metric, we get the value of the path integral to be:
\begin{align}
    \begin{split}
        Z_{\textrm{2DEMSO(3)}}^{g,n}&=\int \mathcal{D}g_{\mu\nu}\mathcal{D}\varphi e^{\int d^2x \sqrt{g}[\varphi R +\frac{6r_0}{L^2}\varphi^{1/2}+2r_0\varphi^{-1/2}] }\times(\text{Topological factor})\\&\exp\Big({{\frac{C_2(R)}{N}}\int d^2x\frac{\sqrt{g}}{2}e^2r_0\varphi^\frac{-3}{2}}\Big)\exp\Big({{\frac{C_2(R')}{N}}\int d^2x\frac{3\sqrt{g}}{2}G_Nr_0\varphi^\frac{-5}{2}}\Big)\,,\\&
        =\Big(\frac{U''(\varphi_0)}{\pi}\Big)^{1/4}\int \mathcal{D}g_{\mu\nu}\exp\Big[\varphi_0(Q)\int d^2x \sqrt{g} R\Big]  \\ &
        \times (\text{Topological factor})\,.\label{A1}
    \end{split}
\end{align}
Recall that we added a boundary counter-term at asymptotic boundaries by adding a defect action, which may cause divergences in the partition function $Z^{g,n}_{2DEMSO(3)}$,but the one-point function which leads to the free energy remains finite.
The path integral in (\ref{A1}) is evaluated as follows \cite{DHoker:1985een,Hatfield:1992rz},
\begin{align}
    \begin{split}
      \bar{Z}:&=  \int \mathcal{D}\varphi \,\exp\Big[{\int d^2x \sqrt{g} \,\underbrace{(\varphi \mathcal{R}-2 U_{Q,j=0}(\varphi))}_{{f(\varphi)}}}\Big]\,,\\ &
      = \int \mathcal{D}\tilde\varphi \,\exp{\int d^2 x \sqrt{g}\Big[\varphi_{0}\mathcal{R}-2U_{Q,j=0}(\varphi_{0})-U''(\varphi_{0})\,\tilde\varphi^2\Big]}\,,\\ &
      =\exp\Big(\int d^2x \sqrt{g}[\varphi_{0}\mathcal{R}-2U_{Q,j=0}(\varphi_0)]\Big)\int \mathcal{D}\tilde \varphi \exp(-\int d^2x \sqrt{g} \,U''(\varphi_0)\,\tilde\varphi^2)\,,\\ &
      =\lim_{p\rightarrow \infty}\exp\Big(\int d^2x \sqrt{g}[\varphi_{0}\mathcal{R}-2U_{Q,j=0}(\varphi_0)]\Big) \Big(\frac{\pi }{U''(\varphi_{0})}\Big)^{p/2}\,.\label{A.4}
    \end{split}
\end{align}
In (\ref{A.4}), there is a divergent term that can be handled by zeta function regularisation. We know
\begin{align}
    \begin{split}
        \prod_{n=1}^{\infty}a\rightarrow\exp \Big(\log(a)\zeta(0)\Big)=a^{-1/2}, \, a=\text{$n$-ind. constant}.\label{A.5m}
    \end{split}
\end{align}
Now using \eqref{A.5m} one can write down the \textit{Zeta-regularised} partition function in \eqref{A.4} as, %\footnote{A simple question may arise: Can one recover the JT gravity results by turning off the second-order fluctuation inside the path-integral? The answer is `NO' because, while path integrating over the dilaton fluctuation, we assume that there is no zero mode. Now, switching off the second-order fluctuation i.e. setting $U''(\varphi_0)=0$ implies that there is a zero mode. Hence, the standard Gaussian integral does not work there. So, to recover the JT-gravity result one should consider up to linear order in fluctuation and do the path integral which gives $\delta(R+2)$.}
\begin{align}
    \begin{split}
     \bar Z\rightarrow   \bar Z_{\zeta}=\Big(\frac{U''(\varphi_0)}{\pi}\Big)^{1/4}\exp\Big(\int d^2x \sqrt{g}[\varphi_{0}\mathcal{R}-2U_{Q,j=0}(\varphi_0)]\Big)
    \end{split}
\end{align}
\textcolor{black}{As mentioned in the main text, we are computing the path-integral over all possible hyperbolic Riemann surfaces, which satisfies the classical e.o.m: $\mathcal{R}-2U'(\varphi)=0$,} along with the classical solution mentioned in (\ref{3.14a}) (also $U(\varphi_0(Q))= 0$ as mentioned in the main text).
%Now to do the path integral over metrics we need to care about the saddle point condition which says $\mathcal{R}=2U'(\varphi_0)<0$.
%%      \int \mathcal{D}g_{\mu\nu}\,\delta(\mathcal{R}-2U'(\varphi_{0})):=\int \hat{\mathcal{D}}g_{\mu\nu}\label{A.8m}
   % \end{split}
%\end{align}}
%The polynomial in $P(\varphi$) is defined in (\ref{3.10}). As we are  looking at $j=0$ sector of SO(3), so $C_2(R')=0$.
 Now we have to perform the path integral over $g_{\mu\nu}.$ This will have two parts. Here we will evaluate the integral over the geodesic boundaries where the extrinsic curvature becomes zero. Therefore, we do not have the Schwarzian term. On the other hand, the extrinsic curvature on asymptotically AdS boundaries produces the Schwarzian, the corresponding contribution to path integral is shown in the main text. So, for the geodesic boundaries, we have the following:
\begin{align}
    \begin{split}
&\int {\mathcal{D}}g_{\mu\nu}\,\exp\Big[\varphi_0(Q)\int d^2x \sqrt{g} \mathcal{R}\Big]\\&=\int {\mathcal{D}}g_{\mu\nu}\Big[\exp\int d^2x \sqrt{g}\mathcal{R}\Big]^{\varphi_0(Q)}\,,\\&
%=\int \mathcal{D}\,g_{\mu\nu}exp\Bigg\{-2\pi\varphi_0(Q)\chi\Bigg\}exp\Bigg\{-\frac{P(\varphi_0(Q))}{L(\varphi_0(Q))}2\pi\chi \Bigg\}\\&
=\int \underbrace{d(\text{Bulk Moduli})}_{\mathcal{V}_{g,1}}\exp\Big[4\pi\chi(\varphi_0(Q)\Big] \,,\\&
= \,\mathcal{V}_{g,1}(b)\exp\Big[4\pi\chi\varphi_0(Q)\Big]\,.\label{A2}
\end{split}
\end{align}
\textcolor{black}{\eqref{A2} has been computed by taking into considering tha tha background solution is asymptotically AdS, hence the surfaces of interest are the hyperbolic Riemann surfaces. Hence, we used Mirzakhani's formula \cite{mirzakhanirecursion} for the volume of moduli space of the hyperbolic Riemann surfaces.}\\
\textcolor{black}{Note that, when we are considering quadratic order fluctuations in dilaton field there must be an back reaction to the spacetime which sets the Ricci scalar $R$ to be negative but non-constant. To see this lets consider the action,
\begin{align}
    \begin{split}
        S&=\int d^2x \sqrt{g}\Big[(\varphi_0+\varphi)\mathcal{R}-2U_{Q}(\varphi_0+\varphi)\Big]+\cdots\\ &
        =\int d^2x \sqrt{g}\Big[(\varphi_0+\varphi)\mathcal{R}-2U_{Q}(\varphi_0)-2U_{Q}'(\varphi_0)\varphi-U_{Q}''(\varphi_0)\varphi^2+\cdots]+\cdots\,.
    \end{split}
\end{align}
Now, the equation of motion of the metric $g_{\mu\nu}$ reads,
\begin{align}
    \begin{split}
        \frac{\delta S}{\delta \varphi}: \mathcal{R}=2U_{Q}'(\varphi_0)+2 U_{Q}''(\varphi_0)\,\varphi:=-\Lambda_Q+2 U_{Q}''(\varphi_0)\,.\varphi\label{A.9a}
    \end{split}
\end{align}
From \eqref{A.9a}, it is clear that the classical solution (which is essentially the saddle point condition in the dilaton path integral) must deviate from $AdS_2$. So, in the gravitational path integral there should be a contribution from the back-reacted geometry. Now the question arises, in which situation one can in-principle ignore the backreaction in the gravitational path integral? In \eqref{A.9a} the extra term $2 U_{Q}''(\varphi_0)\,\varphi\sim \frac{1}{Q^4}$ while in $\Lambda_Q\sim \frac{1}{Q^2}$ and in the large charge limit one can ignore this extra contribution compared to the term ($\Lambda_Q$). }\\
\\
\textbf{A digression on Gaussian integral on curved spacetime:} In curved spacetime, a general Gaussian integral looks like
\begin{align}
    \begin{split}
        Z_{gau.}:=\int \mathcal{D}c_{\mu}\, \exp\Big(-\int d^2 \xi \sqrt{g}\,c^{\mu} \mathcal{O}_{g}c_{\mu}\Big), \,\mu\sim (1,..,d)\,.
    \end{split}
\end{align}
If the operator $\mathcal{O}_{g}$ has some zero modes, that is, $\mathcal{O}_{g}c^{\mu}_{0}=0$, then we have to do the path integral over the zero modes separately as \cite{Hatfield:1992rz},
\begin{align}
    \begin{split}
          Z_{gau.}&=\int \mathcal{D}c_{\mu,0}\int \mathcal{D}c_{\mu}' \,\exp\Big(-\int d^2 \xi \sqrt{g}\,{c'}^{\mu} \mathcal{O}_{g}{c'}_{\mu}\Big)\,,\\ &
          \sim (\int d^2 \xi \sqrt{g})^{d/2}\, \text{det}'(\mathcal{O}_{g})^{-d/2}\,.
    \end{split}
\end{align}
Here, det' denotes the product of all non-zero modes. For our case, $\mathcal{O}_{g}=\hat{\mathds{1}}$ (Identity operator) and $\hat{\mathds{1}}$ do not have any nontrivial zero modes. This implies that the volume term ($\int d^2 \xi \sqrt{g}$) will not be there in (\ref{A.4}). 

\newpage 
\section{Details of the calculation in section \ref{sec6} \label{a2}}

Now we show the details of the computation of $\mathcal{I}_{trumpet}$, which is given by the following integral,\\\\
\textbf{Higher genus contribution to the One-Point function:}
Integral involving the computation of higher genus one point function has the following form:
\begin{align}
    \begin{split}
      &\mathcal{I}^{\text{undef.}}_{\text{trumpet}}\Big|_{g\gg1} =\sum_ge^{4\pi\chi\varphi_0(Q)}\Bigg[\int_0^g b db \, {{\pi^{1/2}}{U''(\varphi_0(Q))}^{1/2}}\,\sqrt{\frac{\varphi_{b,Q}}{2\pi\beta}}\, e^{\frac{-b^2\varphi_{b,Q}}{2\beta}}\frac{4(4\pi^2)^{2g-\frac{3}{2}}}{(2\pi)^{3/2}}\Gamma(2g-\frac{3}{2})\frac{\sinh{b/2}}{b}\exp{(\beta \mu Q)} \\&+\int_g^{\infty} b db\, {{\pi^{1/2}}{U''(\varphi_0(Q))}^{1/2}}\,\sqrt{\frac{\varphi_{b,Q}}{2\pi\beta}}\, e^{\frac{-b^2\varphi_{b,Q}}{2\beta}}{(\frac{3}{2})}^{2g-1}\frac{\Gamma(2g-1)}{\Gamma(6g-2)}\frac{b^{6g-4}}{2\pi}\exp{(\beta \mu Q)}\Bigg]\,, \\&
      =\sum_g \,\,e^{4\pi\chi\varphi_0(Q)}{{\pi^{1/2}}{U''(\varphi_0(Q))}^{1/2}}\sqrt{\frac{\varphi_{b,Q}}{\Delta }}\exp{(\beta \mu Q)}\\ &\Bigg[\frac{4(4\pi^2)^{2g-\frac{3}{2}}}{(2\pi)^{3/2}}\Gamma(2g-\frac{3}{2})\frac{\sqrt{\frac{\pi\beta }{2}}  e^{\frac{\beta }{8 \varphi_{b,Q} }} \left(-2 \text{erfc}\left(\frac{\sqrt{\beta }}{2 \sqrt{2\varphi_{b,Q}} }\right)+\text{erfc}\left(\frac{\beta +2 g \varphi_{b,Q} }{2 \sqrt{2\beta\varphi_{b,Q}} }\right)+\text{erfc}\left(\frac{\beta -2 g \varphi_{b,Q} }{2 \sqrt{2\beta \varphi_{b,Q}}}\right)\right)}{2 \sqrt{\varphi_{b,Q} }}\\&+{2^{3 g-\frac{5}{2}} \left(g^{6 g-3} \left(\frac{g^2 \varphi_{b,Q} }{\beta }\right)^{\frac{3}{2}-3 g} \left(\Gamma \left(3 g-\frac{3}{2},\frac{g^2 \varphi_{b,Q} }{2 \beta }\right)-\Gamma \left(3 g-\frac{3}{2}\right)\right)+\Gamma \left(3 g-\frac{3}{2}\right) \left(\frac{\varphi_{b,Q} }{\beta }\right)^{\frac{3}{2}-3 g}\right)}\Bigg]\,,
     \label{B1}
    \end{split}
\end{align}
where $\Gamma(a,b)$ is the incomplete Gamma function.\footnote{Due to non-zero cutoff $g$.}
Now to calculate $\mathcal{I}^{Trumpet}|_{g>>1}$ we again perform the kernel integration. Using (\ref{3.8}), we get the same delta function. So after the deformation, the integral becomes,
\begin{align}
    \begin{split}
    & \mathcal{I}^{\text{def.}}_{\text{trumpet}}\Big|_{g\gg1}  \\ & =\sum_g{\scriptscriptstyle \,\,U''(\varphi_0(Q))^{1/2}\sqrt{\frac{\varphi_{b,Q}}{\Delta }}\,\Bigg[\frac{\beta}{\Delta f'(\Delta) }\frac{4(4\pi^2)^{2g-\frac{3}{2}}}{(2\pi)^{3/2}}\Gamma(2g-\frac{3}{2})\frac{\sqrt{\frac{\pi \Delta }{2}} e^{\frac{\Delta  }{8 \varphi_{b,Q} }} \left(-2 \text{erfc}\left(\frac{\sqrt{\Delta  }}{2 \sqrt{2\varphi_{b,Q} } }\right)+\text{erfc}\left(\frac{\Delta  +2 g \varphi_{b,Q} }{2 \sqrt{2\Delta\varphi_{b,Q} } }\right)+\text{erfc}\left(\frac{\Delta -2 g \varphi_{b,Q} }{2 \sqrt{2\Delta\varphi_{b,Q}  } }\right)\right)}{2 \sqrt{\varphi_{b,Q} }}}\\&+{\scriptscriptstyle2^{3 g-\frac{5}{2}} \left(g^{6 g-3} \left(\frac{g^2 \varphi_{b,Q} }{\Delta }\right)^{\frac{3}{2}-3 g} \left(\Gamma \left(3 g-\frac{3}{2},\frac{g^2 \varphi_{b,Q} }{2 \Delta  }\right)-\Gamma \left(3 g-\frac{3}{2}\right)\right)+\Gamma \left(3 g-\frac{3}{2}\right) \left(\frac{\varphi_{b,Q} }{\Delta }\right)^{\frac{3}{2}-3 g}\right)}+\Delta \longleftrightarrow \xi\Bigg]e^{4\pi\chi\varphi_0(Q)}\,.\label{B.2}
    \end{split}
\end{align}

\vspace{-0.5cm}
Similarly, we can  calculate the undeformed integral for $g=1,n=1$. For that we have to focus on $\mathcal{V}_{1,1}(b)$.\\

The Weil-Peterson volume at one boundary and one genus is given by,\cite{Saad:2019lba} $$\mathcal{V}_{1,1}(b)=\frac{(b^2+4\pi^2)}{48}\,.$$

\underline{\textit{Lower genus contribution, g=1}}
\begin{align}
\begin{split}
 \mathcal{I}^{\textrm{trumpet}}|_{\textrm{undeformed,\,g=1}}&=\int d\Tilde{h}\int b db \,U''(\varphi_0(Q))^{1/2}\,\sqrt{\frac{\varphi_{b,Q}}{2\pi\beta }}\, e^{\frac{-b^2\varphi_{b,Q}}{2\beta}} \exp(\beta \mu Q)\mathcal{V}_{1,1}(b)e^{-\beta \Tilde{M_0}(Q)} (\textrm{other factors})\,.\\&
 = \int d\Tilde{h}\,\frac{\beta  \left(\beta +2 \pi ^2 \varphi_{b,Q}\right)}{24 \varphi_{b,Q} ^2}U''(\varphi_0(Q))^{1/2}\sqrt{\frac{\varphi_{b,Q}}{\beta }}\,\exp(\beta \mu Q)e^{-\beta \Tilde{M_0}(Q)}(\textrm{other factors})\,.
\end{split}
\end{align}

To calculate the deformed partition function one point function, we need to do the kernel integral as we have done previously.\\

\underline{\textit{Deformed partition function one point function}}
\begin{align}
\begin{split}
\mathcal{I}^{Trumpet}|_{def.,g=1}&=\int_0^\infty d\beta' \frac{\beta e^{-\beta' \Tilde{M_0}(Q)}}{2 \beta'^2 \alpha}\alpha\beta'\Big\{\frac{\delta(\beta'-\Delta)}{|f'(\Delta)|}+\frac{\delta(\beta'-\xi)}{|f'(\xi)|}\Big\}\exp{\frac{(\beta'-\beta)\mu}{\alpha \beta'}} \\ & \hspace{3.5 cm}\times\frac{\beta'  \left(\beta' +2 \pi ^2 \varphi_{b,Q} \right)}{24 \varphi_{b,Q} ^2}\sqrt{\frac{\varphi_{b,Q}}{\beta'}}{U''(\varphi_0(Q))^{1/2}}\,,\\&
=\Bigg[\frac{\beta e^{-\Delta \Tilde{M_0}(Q)}}{\Delta^{3/2} f'(\Delta)}\exp\frac{(\Delta-\beta)\mu}{\alpha \Delta }\frac{\Delta  \left(\Delta  +2 \pi ^2 \varphi \right)}{24 \varphi ^2}+\frac{\beta e^{-\xi \Tilde{M_0}(Q)}}{ \xi^{3/2} f'(\xi)}\exp\frac{(\xi-\beta)\mu}{\alpha \xi }\frac{\xi  \left(\xi +2 \pi ^2 \varphi_{b,Q} \right)}{24 \varphi_{b,Q} ^2}\Bigg]\\ & \hspace{3 cm}\times\sqrt{{\varphi_{b,Q}}}{U''(\varphi_0(Q))^{1/2}}\,.
\end{split}
\end{align}
\bibliography{ref}
\bibliographystyle{utphysmodb}

\end{document}